# Discontinuous-Property Mechanical Metamaterials: A Framework for Discontinuous Lattice Dynamics

Lihong Ao and Aymeric Ramiere*

College of Physics and Optoelectronic Engineering, Shenzhen University, Shenzhen, P.R. China

*ramiere@szu.edu.cn

**Abstract:**

Discontinuity, as the extreme limit of nonlinearity, is common in macroscopic continuum systems. This work introduces discontinuous-property mechanical metamaterials (DPMMs) as a prototype for establishing a geometric-mechanics framework for discontinuous lattice systems. We extend band theory to systems with complementary constraints through geometric regularization and analyze modal discontinuities. By analyzing its static, modal, and transient responses, we validated discontinuous stiffness and Poisson's ratio, alongside intense nonlinear wave mixing and mode conversion. A proof-of-concept experiment demonstrates the practical feasibility of implementing DPMMs. This work characterizes discontinuous lattice systems through the intuitive DPMM concept and provides a practical framework to harness sharp transitions in mechanical metamaterials for nonlinear wave manipulation.

## I. Introduction

Mechanical metamaterials are a class of architected structures that derive their exceptional properties, such as negative Poisson's ratio, negative moduli, and tunable band gaps, from engineered microstructures rather than from the intrinsic properties of their constituent materials [1–4]. Unlike conventional materials, in which macroscopic behavior is determined by chemical composition and atomic-scale bonding, mechanical metamaterials rely on the rational design of unit cells, enabling properties rarely observed in nature [5,6]. Examples include rotating hinge units [7,8], open-cell foams [9], and Kresling origami structures [10,11]. These architectures have enabled applications in aerospace, soft robotics, and vibration isolation [12–15], demonstrating strong innovation potential.

In parallel, nonlinearity and interparticle interactions in condensed matter physics drives emergent phenomena [16]. From the optical Kerr effect [17,18], to Kondo effect [19,20], and to exotic nonlinear responses in symmetry-broken materials, nonlinearity lies at the heart of many striking phenomena. Unlike linear systems, where the response is proportional to the excitation, nonlinear systems exhibit thresholds, multi-stability, hysteresis, and qualitatively new behavior as control parameters vary [21,22]. In cases where interactions are weak, linear band theory often provides a sufficient description [23,24]. Conversely, if infinitely many particles interact, nonlinear effects become strongly localized and concentrates at singularities, leading to fundamentally different physics that cannot be described by analytic nonlinearity using finite-order approximations. This case has not found in the real microscopic world, but it is made possible by macroscopic metamaterials under a continuum assumption.

The intersection of metamaterials and nonlinear physics suggests the possibility of designing a periodic system whose unit cells interact through macroscopic mechanical contact, thereby concentrating nonlinearity at geometric singularities. These unit cells would exhibit a state-dependent, infinitely sensitive nonlinear change in their mechanical properties. In such a mechanical metamaterial, a discontinuity occurs when an arbitrarily small applied strain does not produce an arbitrarily small change in mechanical properties. This would represent a fundamentally new form of nonlinearity, in which the nonlinearity is concentrated at a nowhere dense discontinuity set. The collective behavior of many interacting unit cells with discontinuous properties is the main subject of our research.

So far, state-dependent property discontinuities in metamaterials have been investigated through individual examples in isolation. For instance, origami metamaterials exhibit a discontinuous Poisson’s-ratio – strain relation [7,25]. Discontinuity-like stiffness – strain relations have also been observed in contact-based designs [12,26,27], its effect on transport properties has been confirmed [28]. Superconducting metamaterials exhibit a discontinuous change in permeability with the electric field [29]. Furthermore, strong nonlinearities have been reported in phononic metamaterials by controlling contact between scatterers to tune acoustic properties and topological phonon states [30,31]. Such discontinuous behavior has broad potential applications in mechanically programmable devices [32–34]. Theoretically, such behaviors are related to the dynamics of non-smooth systems that have been studied in the context of bifurcation theory and basic impact oscillators [35–39]. However, these general theories are not convenient for classifying such systems or guiding their design.

To address these gaps, we introduce the concept of Discontinuous Property Mechanical Metamaterials (DPMMs). A DPMM is a macroscopic, periodic structure composed of unit cells with a state-dependent discontinuous property change. Here, we propose a unified framework for DPMMs that connects the concepts of non-smooth dynamics and lattice systems and extends band theory to cases with complementary constraints by a geometric regularization. We present an example and use numerical simulations to show how the property discontinuity affects the static, modal, and dynamic characteristics. A DPMM prototype is fabricated to demonstrate that strong nonlinearity can be achieved experimentally. This work bridges mechanical contact in metamaterials with nonlinear dynamics in lattice systems, establishing a foundation for studying discontinuities in mechanical metamaterials.

## II. Mechanics of Discontinuity Lattice System

DPMMs are based on periodic unit cells designed to have mechanical contacts. We start by establishing a theoretical framework for contact-induced discontinuities, which are more singular than conventional nonlinearities. The framework includes the solid-mechanics equations with contact forces and the behavior of the band structure across the discontinuity.

### A. Concept of contact discontinuity

Contact in continuum mechanics is modeled by the unilateral Signorini conditions in Eq. (2). In this ideal model, contact pressure is a constraint reaction, not a single-valued constitutive function of gap. Contact activation can change the admissible incremental displacement. Specified interface law and activation geometry results in a finite jump in tangent stiffness, makes the contact pressure at least a non-smooth function of the gap. Some models like augmented-Lagrangian law even regard the contact pressure itself as a discontinuous function [40]. The DPMM architecture considered here uses contact with external walls to modify its effective stiffness and lateral response. Figure 1(a) illustrates the idealized bilinear response. Near an activation threshold, mechanical waves may switch the contact state and thereby undergo nonlinear mixing; the magnitude of this effect depends on the interface model, preload, and excitation amplitude.

The concept of DPMMs is proposed to introduce discontinuous systems. We construct a dynamical system with discontinuous external forces (non-smooth potential) under parameters described by a vector $\boldsymbol{P}$, and the evolution of its phase (spatial-momentum) coordinate $\boldsymbol{q}$ can be generally described by the following equation[41]:

$$\dot{\boldsymbol{q}} = f\big(\boldsymbol{q}, \boldsymbol{P}, V_{non-smooth}(\boldsymbol{q}, \boldsymbol{P})\big) \tag{1}$$

Here, $\boldsymbol{P}$ includes various design and manipulation parameters, such as external strain. $V_{non-smooth}$ is the non-smooth contact potential, which is also known as the complementary potential in contact mechanics[42]. With $N$ contact pairs, the contact gap function $g_i$ and contact pressure $p_i$ satisfies the Signorini conditions:

$$g_i(\boldsymbol{q},\boldsymbol{P}) \geq 0, p_i \geq 0, p_i g_i(\boldsymbol{q},\boldsymbol{P}) = 0 \quad (i = 1,2,\dots,N, N \geq 1) \tag{2}$$

The contact gap function $g_i$ means the minimum distance between two contact surfaces and is only related to the position coordinates[43]. This condition states that solids do not penetrate each other, and all deformation after contact arises from the deformation of the solids themselves. For surface-to-surface contact, the stiffness undergoes a discontinuous change at $g_i = 0$.

Here, it is extended to the generalized parameter coordinates $(\boldsymbol{q},\boldsymbol{P})$ in generalized parameter space $\mathcal{P}$, which is a Banach manifold. For multiple contact pairs, define discontinuous parameter set $\Sigma_i \equiv \{(\boldsymbol{q},\boldsymbol{P}) | g_i(\boldsymbol{q},\boldsymbol{P}) = 0\}$. $\Sigma = \cup_i \Sigma_i = \{(\boldsymbol{q},\boldsymbol{P}) | V(\boldsymbol{q},\boldsymbol{P})\ is\ nonsmooth\}$ is called the discontinuity set (DS). Normal contact pressure $p_i(g_i)$ is at least non-smooth at the DS. The gradient of the gap function at the contact surface is non-zero, that is the surjective derivative condition $dg_i(\boldsymbol{q},\boldsymbol{P}) \neq 0$. For continuum mechanics with infinite-dimension, the contact surfaces have definite directions, $\ker\left(dg_i(\boldsymbol{q},\boldsymbol{P})\right)$ has a topological complement space within $\mathcal{P}$ (splitting condition). By the implicit function theorem with the above conditions, $\Sigma$ is a codimension-one smooth submanifold embedded in $\mathcal{P}$. Intersections of several $\Sigma_i$ have codimension equal to the rank of the active gap gradients, provided the corresponding transversality conditions hold. Such a geometric picture depicts that all the singular properties reside in these thin "surfaces" in the parameter space, while perturbations occurring elsewhere are nothing special.

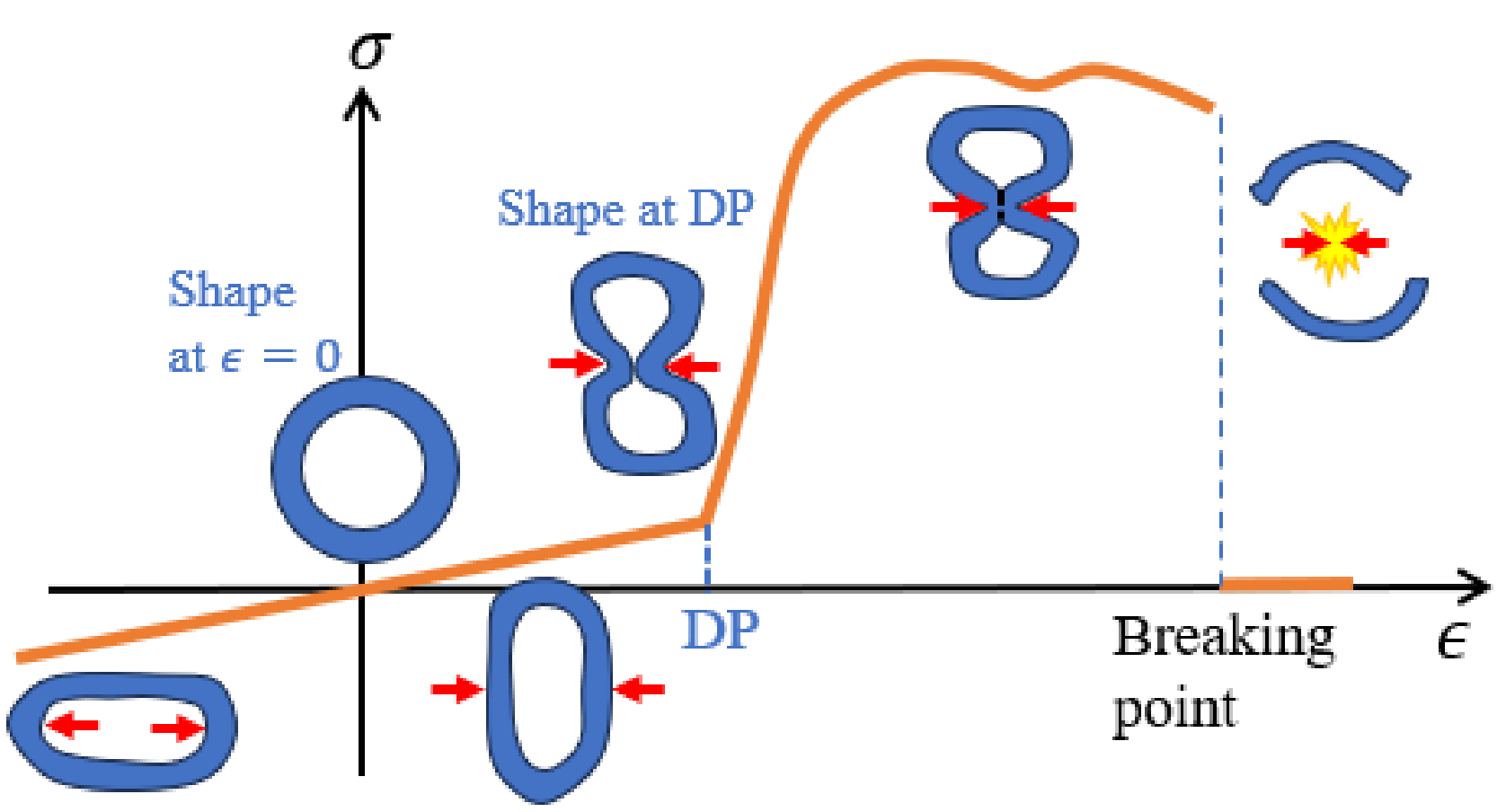

**Figure 1.** (a) Schematic of contact stiffness discontinuity (for illustration only. In reality, contact discontinuity imposes requirements on the contact surface geometry).

## B. Elastic solid equation with surface-to-surface contact

The bulk material of DPMM in this work is considered to be an ordinary isotropic solid in classical continuum mechanics. The motion state of DPMM can be determined by the displacement vector of DPMM $|\boldsymbol{u}(\boldsymbol{r},t)\rangle$. For isotropic elastic case, using the Navier equation in solid mechanics:

$$\rho\partial_t{}^2|\boldsymbol{u}\rangle = (\kappa\boldsymbol{\nabla}^2 + (\kappa+\lambda)\boldsymbol{\nabla}\boldsymbol{\nabla}\cdot)|\boldsymbol{u}\rangle + |\boldsymbol{f}_c\rangle \tag{3}$$

where $\kappa, \lambda$ are the Lamé coefficients. $|\boldsymbol{u}\rangle$ is a function of the spatial coordinate $\boldsymbol{r}$, and its spatial domain of definition is the smooth domain $\Omega \subset \mathbb{R}^3$ occupied by the DPMM. Strain tensor $\hat{\boldsymbol{\epsilon}} = \frac{1}{2}(\boldsymbol{\nabla}|\boldsymbol{u}\rangle + \langle\boldsymbol{u}|\boldsymbol{\nabla})$ and stress tensor $\hat{\boldsymbol{\sigma}} = 2\kappa\hat{\boldsymbol{\epsilon}} + \lambda(\boldsymbol{\nabla}\cdot|\boldsymbol{u}\rangle)\boldsymbol{\delta}$ can be derived from the displacement field, with $\boldsymbol{\delta}$ the Kronecker symbol. $|\boldsymbol{f}_c\rangle$ is the external force acting on the system. Here, we only consider the contact force. It includes the normal pressure and may include friction. The contact pressure applied on a DPMM in contact with a surface of local normal basis $|\boldsymbol{n}\rangle$ is given by $p_n = \langle\boldsymbol{f}_c|\boldsymbol{n}\rangle = \langle\boldsymbol{n}|\hat{\boldsymbol{\sigma}}\boldsymbol{n}\rangle$. This pressure generates a contact friction with magnitude $p_t \le \mu p_n$, where $\mu$ is the friction coefficient. Gap function $g$ is determined by $|\boldsymbol{u}\rangle$. Therefore, $|\boldsymbol{f}_c\rangle$ is not explicit and introduces discontinuity in stiffness when contact behavior is present. From the perspective of non-equilibrium phase transition, the cell in the separated state ($g > 0$) and the contact state ($g \le 0$) represent distinct mechanical regimes. The transition between them can be viewed as a switching process that breaks the discrete $\mathbb{Z}_2$ symmetry of the DPMM. This is analogous to a first-order-like transition in the state space, where the contact constraint imposes a complementary condition.

Equation (3) is used to solve the transient behavior of DPMM even if the stress-strain relation $\sigma(\epsilon)$ has high-order nonlinearity, such as a perturbation at the discontinuity set (DS) and dynamic wave coupling. The static properties of DPMM, such as $\sigma(\epsilon)$, are solved in static case ($\partial_t|\boldsymbol{u}\rangle = 0$). Different boundary conditions are applied in different cases in the following simulations, and will be stated in the result description.

Similar to Eq. (3) for the spatial domain, the frequency domain is studied with the following equation:

$$(\kappa\boldsymbol{\nabla}^2 + (\kappa+\lambda)\boldsymbol{\nabla}\boldsymbol{\nabla}\cdot)|\boldsymbol{u}_m\rangle + |\boldsymbol{f}_c\rangle = -\rho\omega_m{}^2|\boldsymbol{u}_m\rangle \tag{4}$$

This form is valid almost everywhere except at the DS. For a unit cell in a periodic system without friction or contact state transition, Eq. (4) satisfies the complementary condition in Eq. (2), and the Bloch periodic condition

$$|\boldsymbol{u}_m(\boldsymbol{r}_{\mathrm{b1}})\rangle = |\boldsymbol{u}_m(\boldsymbol{r}_{\mathrm{b0}})\rangle \mathrm{e}^{-\mathrm{i}\boldsymbol{k}(\boldsymbol{r}_{\mathrm{b1}}-\boldsymbol{r}_{\mathrm{b0}})} \tag{5}$$

The solution of Eq. (4) is composed of normalized eigenstates $|\boldsymbol{u}_{m,\boldsymbol{k}}\rangle$ with eigenvalues $\omega_{m,\boldsymbol{k}}$. The subscript $m$ is the mode index, with the number of modes in a continuous medium being infinite. $\boldsymbol{k}$ is the wave vector. $\boldsymbol{r}_{\mathrm{b0}}$ and $\boldsymbol{r}_{\mathrm{b1}}$ are the coordinates of the source and destination boundaries where periodic boundary conditions are applied. These two boundaries are parallel to each other and are spaced at a lattice constant along the periodic direction. Including the effect of contact in Eq. (4) yields a complementary eigenvalue problem[44,45]. The complementary conditions restrict the existence of some modes.

Contact does not necessarily break translational symmetry: a uniformly periodic contact pattern can preserve Bloch periodicity, whereas nonuniform contact pattern can reduce the symmetry and invalidate the simple Bloch condition. Nevertheless, we can still consider the contact phenomenon from the perspective of the phonon field. If the Hermite (frictionless and undamped) mechanical energy is extended into harmonic terms with different $\boldsymbol{k}$ in non-contact region, the mechanical energy of the phonon field cannot be expressed by finite-order Hermitian polynomials. Suppose to the contrary, if its mechanical energy is finite order, we decompose the modes and define:

$$a_{m,\boldsymbol{k}} = \int d\boldsymbol{r} e^{-ikr}\left(\sqrt{\frac{\rho\omega_{m,\boldsymbol{k}}}{2}}\left|\boldsymbol{u}_{m,\boldsymbol{k}}\right\rangle + \frac{i}{\sqrt{2\rho\omega_{m,\boldsymbol{k}}}}\left|\dot{\boldsymbol{u}}_{m,\boldsymbol{k}}\right\rangle\right)$$

$$a^*_{m,\boldsymbol{k}} = \int d\boldsymbol{r} e^{-ikr}\left(\sqrt{\frac{\rho\omega_{m,\boldsymbol{k}}}{2}}\left|\boldsymbol{u}_{m,\boldsymbol{k}}\right\rangle - \frac{i}{\sqrt{2\rho\omega_{m,\boldsymbol{k}}}}\left|\dot{\boldsymbol{u}}_{m,\boldsymbol{k}}\right\rangle\right) \tag{6}$$

where $\left|\dot{\boldsymbol{u}}_{m,\boldsymbol{k}}\right\rangle = \partial_t\left|\boldsymbol{u}_{m,\boldsymbol{k}}\right\rangle$ is the modal velocity field. Let $V_{\text{nonlinear}}$ represent the nonlinear component of the non-smooth contact potential and write the total mechanical energy from the perspective of modal superposition:

$$\begin{aligned}\mathcal{H} &= \frac{1}{2}\left(\rho\langle\dot{\boldsymbol{u}}|\dot{\boldsymbol{u}}\rangle + \lambda\langle\boldsymbol{\nabla}\cdot|\boldsymbol{u}\rangle^2 + 2\mu|\boldsymbol{\epsilon}|^2\right) + V_{\text{nonlinear}} \\ &= \sum_{\alpha,\beta=1}^{I}\sum\nolimits_{m,k} c_{\alpha,\beta}\left(a^*_{m,\boldsymbol{k}}\right)^{\alpha} {a_{m,\boldsymbol{k}}}^{\beta} + \mathcal{H}_0\end{aligned} \tag{7}$$

where $I$ is a finite integer. The coefficient $c_{\alpha,\beta} = c^*_{\alpha,\beta}$, $c_{1,1} = \omega_{m,\boldsymbol{k}}$. $\mathcal{H}_0$ is a constant. $\omega_{m,\boldsymbol{k}}\left(a^{\dagger}_{m,\boldsymbol{k}}a_{m,\boldsymbol{k}} + \frac{1}{2}\right)$ is the energy of a linear oscillator, while other terms come from the nonlinearity. In Eq. (7), Bloch modes $\left|\boldsymbol{u}_{m,\boldsymbol{k}}\right\rangle$ can be a $C^{\infty}$-smooth in the position space. $V_{\text{nonlinear}}$ as its finite linear combination should be also $C^{\infty}$-smooth in this case. The contradiction arises that $V_{\text{nonlinear}}$ is actually a non-smooth one. Thus, when strictly expanding an Hermite non-smooth contact potential, we cannot obtain any simplified form, the only possibility of its linear modal expansion is using the general form (infinite-order) of a Hermitian Hamiltonian[46,47]:

$$\mathcal{H} = \sum_{\alpha,\beta=1}^{\infty}\sum\nolimits_{m,k} c_{\alpha,\beta}\left(a^*_{m,\boldsymbol{k}}\right)^{\alpha} {a_{m,\boldsymbol{k}}}^{\beta} + \mathcal{H}_0, \qquad (\boldsymbol{q},\boldsymbol{P}) \in \mathcal{R}\backslash\Sigma \tag{8}$$

Eq. (8) gives the mechanical energy of DPMM without friction and damping, which can be interpreted as a global polynomial approximation on a compact amplitude domain. Its gradient converging locally uniformly only outside $\Sigma$, and converging globally in the $L^p$ or weak sense[48]. The order $\alpha + \beta$ represents the orders of nonlinearity, that is, the number of acting internal finite-order polynomials. For instance, optical frequency doubling and Kerr effect are third-order nonlinear (two-photon interaction). In the same way, the interaction of $I$ particles produces $2I - 1$ order nonlinearity. DPMMs use phonon interactions with the propagation of classical mechanical waves where contact nonlinearity corresponds to interaction terms of arbitrary high order because terms of all orders may contribute to the dynamics. It stems from the cooperative motion of infinitely many degrees of freedom at the contact interface under the continuum limit. This non-analyticity implies that conventional perturbation theory fails near the DS, and the system must be treated piecewise.

Consequently, even small mechanical waves can experience large scattering, potentially leading to strong wave mixing. Even with a finite-order approximation, many higher-order terms cannot be ignored, leading to significant nonlinearity near the DS and possible strong phonon-phonon coupling. This implies that near the DS, the nonlinearity is so strong that perturbation theory in the usual sense fails, but the system remains deterministic and can be simulated directly. These conditions make DPMMs highly sensitive to external factors and may produce other non-smooth dynamics phenomena in suitable parameter regimes. (see *Appendix A* for background).

## C. Discontinuous band structure

Under the geometric regularization of II.A, the properties on besides the DS can be linearized, and be described by conventional band theory. The only thing that needs to be solved is to incorporate DS into the band structure. Previous work [49] introduced the concept of "non-smooth modes", which refers to a continuous orbit family of periodic solutions, whose period is related to the amplitude. These orbits are directly constructed by solving a set of nonlinear equations that include the number of collisions, without the need for regularization of discontinuities. However, these non-smooth modes do not have orthogonality, so they cannot be simply linearly superimposed and the calculation is complex. In this work, we do not employ this non-smooth modal analysis. Instead, we still describe the modal behavior of the discontinuous system based on the linear parametric eigenvalue problem but perform regularization on the geometry of the band structure.

First, let's extend the parameter $\boldsymbol{k}$ in Eq. (3) to the general parameter $\boldsymbol{P}$, which not only includes the wave vector, but also the manipulation parameters and design parameters. These parameter vectors constitute the parameter space $\mathcal{P}$. Then, note $\mathcal{L} \coloneqq -\frac{1}{\rho}(\kappa\boldsymbol{\nabla}^2 + (\kappa+\lambda)\boldsymbol{\nabla}\boldsymbol{\nabla}\,\cdot)$ and $|\boldsymbol{f}_c\rangle = -\mathcal{V}(\boldsymbol{r})|\boldsymbol{u}\rangle$, such that Eq. (4) becomes a complementary eigenvalue problem $(\mathcal{L}+\mathcal{V})|\boldsymbol{u}_m\rangle = {\omega_m}^2|\boldsymbol{u}_m\rangle$, where $\mathcal{V}$ is an operator that is non-zero only on contact surface $\mathcal{G}$. For frictionless contact, it acts only on the normal component:

$$\mathcal{V}(\boldsymbol{r}) = \begin{cases} \frac{k_n}{\rho}|\boldsymbol{n}\rangle\langle\boldsymbol{n}| \times \delta(r-\mathcal{G}), & contact \\ 0, & non-contact \end{cases} \tag{9}$$

The full contact set $\Sigma \subset \mathcal{R}$ defined in Sec. II.A need not project smoothly onto $\mathcal{P}$. Let $\boldsymbol{q}_{\mathrm{nc}}(\boldsymbol{P})$ be the continuously differentiable continuation of the non-contact equilibrium near a candidate switch. Define the parameter-space gap $\hat{g}(\boldsymbol{P}) = g(\boldsymbol{q}_{\mathrm{nc}}(\boldsymbol{P}), \boldsymbol{P})$ and parametric DS $\hat{\Sigma} = \hat{g}^{-1}(0)$. If $d\hat{g}(\boldsymbol{P}_*) \neq 0$ at $\boldsymbol{P}_* \in \hat{\Sigma}$, an admissible parameter perturbation opens or closes the gap at first order, and the finite-dimensional implicit-function theorem makes $\hat{\Sigma}$ a local smooth hypersurface of $\mathcal{P}$. Within each smooth region $\mathcal{P} \setminus \hat{\Sigma}$, the tangent problem has eigenpairs $(\omega_{m,\boldsymbol{P}}, |\boldsymbol{u}_{m,\boldsymbol{P}}\rangle)$ and modal bundle $\mathcal{M}_m$. Thus only the two one-sided limits need be compared.

For comparison, in conventional band structure theory, Kato-Rellich theorem[50,51] tells that: for locally continuous $\boldsymbol{P}$, $\omega_m$ are also locally continuous, while $|\boldsymbol{u}_m\rangle$ are discontinuous only where the eigenvalues are degenerate, and analytic elsewhere. For instance, the Dirac points [52–54], Weyl points[55–57], and nodal lines[58–60] are places where the eigenvalues are degenerate. They are singularities of Berry curvature, meaning that the eigenvectors are discontinuous, but the eigenvalues remain continuous. Outside these singularities of Berry curvature, the eigenvalues and eigenstates are both analytic. However, when we introduce discontinuous external forces $|\boldsymbol{f}_c\rangle$ (or contact potential $V_{non-smooth}$), standard Kato perturbation theory cannot be applied directly across $\hat{\Sigma}$ because contact activation changes the tangent form discontinuously. For finite activated contact stiffness, the two one-sided tangent operators differ by a nonnegative contact form. We show that, a contact-observable simple mode has an eigenvalue jump and eigenstates jump, an observable isolated cluster has a spectral-sum jump. The reason is discussed below.

We define the domain of a unit cell as $\Omega$, note the contact surface as $\mathcal{G} \subset \Omega$ with normal basis $|\boldsymbol{n}\rangle$, and let $|\boldsymbol{v}\rangle$ be a test function (a virtual displacement field) to construct the following weak form for bilinear stiffness and mass:

$$a\,(|\boldsymbol{u}_m\rangle\,, |\boldsymbol{v}\rangle) = \int_\Omega \widehat{\boldsymbol{\sigma}}\colon \hat{\epsilon}\, d\boldsymbol{r}\,, \qquad b(|\boldsymbol{u}_m\rangle\,, |\boldsymbol{v}\rangle) = \int_\Omega \rho\langle\boldsymbol{u}_m|\boldsymbol{v}\rangle\, d\boldsymbol{r} > 0 \tag{10}$$

The linear differential operator $\mathcal{L} := -\frac{1}{\rho}(\kappa\boldsymbol{\nabla}^2 + (\kappa+\lambda)\boldsymbol{\nabla}\boldsymbol{\nabla}\cdot)$ satisfies $\langle\mathcal{L}|\boldsymbol{u}_m\rangle|\boldsymbol{v}\rangle = a\,(|\boldsymbol{u}_m\rangle\,,|\boldsymbol{v}\rangle)$ and has standard elliptic regularity (see *Appendix B*).

Fix $\boldsymbol{P}_* \in \hat{\Sigma}$, with $\hat{g}(\boldsymbol{P}) > 0$ on the non-contact side and $\hat{g}(\boldsymbol{P}) < 0$ on the active-contact side. Pull nearby cells back by uniformly bi-Lipschitz maps to a bounded Lipschitz reference cell $\Omega_*$. After the Bloch gauge transformation and mass normalization, all tangent forms share the fixed compact embedding $H^1_{\mathrm{B}}(\Omega_*;\mathbb{C}^3) \hookrightarrow L^2(\Omega_*;\mathbb{C}^3)$, where the subscript B denotes the essential and Bloch boundary conditions. The displacement space is therefore infinite-dimensional, whereas every isolated spectral subspace below is finite-dimensional.

In the physical setting considered here, the bulk material and cell geometry remain unchanged at the moment of contact, while activation changes the interface normal stiffness. Accordingly, assume that the pulled-back bulk elastic and mass forms are form-norm continuous on each side of $\hat{\Sigma}$ and have the same two one-sided limits, with uniform positive density bounds, uniform strong ellipticity (*Appendix B*), a uniform Korn inequality, and linear stability of the selected equilibrium. These assumptions exclude buckling, loss of material stability, geometric degeneration, and unconstrained rigid-body motion.

The Signorini admissible set is a cone, so modes are defined by the tangent operator at a fixed equilibrium and active set. Let $a_-$ be the non-contact tangent form at $\boldsymbol{P}_*$. On an active frictionless patch $\mathcal{G}_{\mathrm{act}}$ of positive surface measure, assume that the finite normal tangent stiffness $k_n$ is essentially bounded above and below by positive constants. With $\gamma$ denoting the trace operator, define

$$\mathfrak{C}_{\mathrm{act}}(|\boldsymbol{u}\rangle,|\boldsymbol{v}\rangle) = \int_{\mathcal{G}_{\mathrm{act}}} k_n\,\langle\boldsymbol{n}|\gamma\boldsymbol{u}\rangle\overline{\langle\boldsymbol{n}|\gamma\boldsymbol{v}\rangle}dS, a_+ = a_- + \mathfrak{C}_{\mathrm{act}}$$

$$a_\pm(|\boldsymbol{u}_j^\pm\rangle,|\boldsymbol{v}\rangle) = \left(\omega_j^\pm\right)^2 b(|\boldsymbol{u}_j^\pm\rangle,|\boldsymbol{v}\rangle), \forall|\boldsymbol{v}\rangle \in H^1_{\mathrm{B}} \tag{11}$$

Strong ellipticity (*Appendix B*), Korn's inequality, and the trace theorem make $a_\pm$ closed and lower-bounded on the common form domain. Compact embedding then gives self-adjoint operators $\mathcal{L}_-,\mathcal{L}_+$ with compact resolvent. Their spectra are discrete and their eigenspaces are finite-dimensional. Strong ellipticity and compact embedding give discreteness; the one-sided form-norm continuity gives the one-sided limits. One-sided form-norm continuity implies norm-resolvent convergence $\mathcal{L}(\boldsymbol{P}) \to \mathcal{L}_-$ as $\hat{g}(\boldsymbol{P}) \to 0^+$, and $\mathcal{L}(\boldsymbol{P}) \to \mathcal{L}_+$ as $\hat{g}(\boldsymbol{P}) \to 0^-$. Hence every isolated eigenvalue cluster has one-sided limits. For a simple cluster, a $b$-normalized eigenstate representing it has one-sided limits after independent phase choices.

1. *Simple isolated mode*. This is the most common case in reality. Assume, there is a simple isolated eigenvalue along $a_\tau = a_- + \tau\mathfrak{C}_{\mathrm{act}}$, bounded $\mathfrak{C}_{\mathrm{act}} \geq 0$, and $\mathfrak{C}_{\mathrm{act}}(|\boldsymbol{u}\rangle,|\boldsymbol{u}\rangle) > 0$. These express a stable equilibrium, finite interface stiffness, and a mode that moves normally on the active patch. Compare the endpoints through $a_\tau = a_- + \tau\mathfrak{C}_{\text{act}}, 0 \le \tau \le 1$. The selected eigenvalue remains simple and separated from neighboring bands for every $\tau$, and choose a continuous $b$-normalized eigenstate $|\boldsymbol{u}_m(\tau)\rangle$. Contact observability means $\mathfrak{C}_{\text{act}}(|\boldsymbol{u}_m(0)\rangle, |\boldsymbol{u}_m(0)\rangle) > 0$; physically, this mode moves normally on the active patch. The Hellmann-Feynman identity and continuity at $\tau = 0$ give

$$\frac{d}{d\tau}(\omega_m(\tau))^2 = \mathfrak{C}_{\text{act}}(|\boldsymbol{u}_m(\tau)\rangle, |\boldsymbol{u}_m(\tau)\rangle) \ge 0$$

$$(\omega_m^+)^2 - (\omega_m^-)^2 = \int_0^1 \mathfrak{C}_{\text{act}}(|\boldsymbol{u}_m(\tau)\rangle, |\boldsymbol{u}_m(\tau)\rangle)d\tau > 0 \tag{12}$$

Eq. (12) demonstrates the discontinuity of eigenvalues, next we discuss the discontinuity of eigenstates. Suppose, contrary to the desired eigenstate discontinuity, that $|\boldsymbol{u}_m^+\rangle = e^{i\varphi}|\boldsymbol{u}_m^-\rangle$ for some $\varphi \in \mathbb{R}$. Multiplying $|\boldsymbol{u}_m^+\rangle$ by $e^{-i\varphi}$ leaves the eigenproblem unchanged and gives a common normalized state $|\boldsymbol{u}\rangle$. Subtracting the two endpoint weak equations yields

$$\mathfrak{C}_{\text{act}}(|\boldsymbol{u}\rangle, |\boldsymbol{v}\rangle) = [(\omega_m^+)^2 - (\omega_m^-)^2] b(|\boldsymbol{u}\rangle, |\boldsymbol{v}\rangle), \forall |\boldsymbol{v}\rangle \in H_{\text{B}}^1 \tag{13}$$

For every $|\boldsymbol{v}\rangle \in C_c^\infty(\Omega_*; \mathbb{C}^3)$, the contact trace vanishes. The strict eigenvalue inequality and density in $L^2(\Omega_*; \mathbb{C}^3)$ then force $|\boldsymbol{u}\rangle = 0$, contradicting $b(|\boldsymbol{u}\rangle, |\boldsymbol{u}\rangle) = 1$. Thus no $U(1)$ phase relates the two one-sided eigenstates. Since $U(1)$ is compact, their phase-minimized mass distance is strictly positive.

2. *Multiple isolated mode*. Set $\nu = \dim \mathcal{M}(\tau) \ge 2$. Assume that the selected isolated cluster remains separated from the rest along the same interpolation. Count its frequencies with multiplicity, and choose a $b$-orthonormal eigenbasis of the cluster, denoted by $|\boldsymbol{u}_j(\tau)\rangle$. Contact observability means that at least one vector of the non-contact modal subspace has positive contact energy. The trace Hellmann-Feynman identity is

$$\frac{d}{d\tau}\sum_{j=1}^{\nu}\left(\omega_j(\tau)\right)^2 = \sum_{j=1}^{\nu} \mathfrak{C}_{\text{act}}(|\boldsymbol{u}_j(\tau)\rangle, |\boldsymbol{u}_j(\tau)\rangle) \ge 0 \tag{14}$$

Continuity at $\tau = 0$ makes the right-hand side positive on a nonzero interval. Therefore the basis-independent cluster sum has the strict endpoint jump

$$\sum_{j=1}^{\nu}\left(\omega_j^+\right)^2 - \sum_{j=1}^{\nu}\left(\omega_j^-\right)^2 > 0 \tag{15}$$

For $\nu > 1$, individual branches need not all jump and their eigenstates have $U(\nu)$ basis freedom. If the two endpoint modal subspaces coincided, both endpoint operators would leave their common subspace invariant. For any vector in it, their difference would therefore lie in the same subspace; testing against compactly supported fields would force that difference, and hence the contact form, to vanish. This contradicts contact observability. Thus the invariant conclusion in the degenerate case is the distinct modal subspaces together with the strict spectral-sum jump; no branchwise eigenstate claim is made.

3. *Genericity.* On a real-analytic piece of $\hat{\Sigma}$, assume that the pulled-back material data and boundary maps depend real-analyticly on $\boldsymbol{P}$. A local finite-dimensional Hermitian reduction of an isolated cluster then has a real-analytic characteristic polynomial. If its discriminant is not identically zero, the repeated-eigenvalue set is a real-analytic subset of the finite-dimensional switching manifold with empty interior and measure zero; simple modes are therefore generic. If the discriminant vanishes identically, multiplicity is enforced by exact symmetry, repeated substructure, or a restricted design family, and the multiple-cluster statement above is the appropriate one.

4. *Rigid bodies contact*. The argument concerns finite post-contact tangent stiffness, as realized by a compliant interface or by finite elasticity of the contacting bodies. Extending the same family to $\tau \geq 0$, ideal hard contact is the monotone limit

$$a_\tau = a_- + \tau \mathfrak{C}_{\text{act}}, \tau \to \infty, \{|\boldsymbol{u}\rangle \in H^1_{\text{B}}: \langle \boldsymbol{n}|\gamma \boldsymbol{u}\rangle_{\mathcal{G}_{\text{act}}} = 0\} \tag{16}$$

More precisely, let $H_\infty = \{|\boldsymbol{u}\rangle \in H^1_B: \gamma_n|\boldsymbol{u}\rangle = 0 \ on \ \mathcal{G}_{\text{act}}\}$; the monotone forms $a_\tau = a_- + \tau \mathfrak{C}_{\text{act}}$ converge to the closed form on $H_\infty$ as $\tau \to \infty$, and the discrete eigenvalues converge by the monotone-form (Mosco) theorem.

$$|(\omega_m^+)^2 - (\omega_m^-)^2| > 0, \inf_{\varphi \in \mathbb{R}} \parallel |\boldsymbol{u}_m^+\rangle - e^{i\varphi}|\boldsymbol{u}_m^-\rangle \parallel_b > 0 \tag{17}$$

From this, it can be understood that discontinuous band structure is formed by a base manifold divided into smooth connected partitions $\mathcal{P}_j (j = 1,2, \dots)$ by its submanifolds $\hat{\Sigma}$. The eigenvalues are continuous within each partition. Eigenvalues are continuous within each partition and smooth away from degeneracies. On the switching set, a contact-observable simple mode has distinct one-sided eigenvalues and phase-inequivalent eigenstates; an isolated multiple cluster has a strict one-sided jump in its spectral sum, while individual branches are not canonically comparable. When the local discriminant is not identically zero, simple modes are generic; otherwise the multiple-cluster statement applies within the symmetry-restricted family. By contrast, in conventional band structure, eigenvalues are continuous at the degeneracy, and the eigenvectors are not necessarily continuous at the degeneracy but are analytic elsewhere. Such regularization enables a natural generalization of the geometry properties of conventional band structures. In regions where a band is smooth, local properties such as Berry connection $\mathcal{A}_m(\boldsymbol{P}) = i\langle \boldsymbol{u}_m(\boldsymbol{P})|d\boldsymbol{u}_m(\boldsymbol{P})\rangle$ and Berry curvature $\Omega_m(\boldsymbol{P}) = d\mathcal{A}_m(\boldsymbol{P}) = \frac{1}{2}\sum_{\alpha,\beta}(\partial_\alpha \mathcal{A}_{m,\beta} - \partial_\beta \mathcal{A}_{m,\alpha}) d\boldsymbol{P}^\alpha \wedge d\boldsymbol{P}^\beta$ can still be defined like that in traditional band theory. Mechanical contact therefore defines a spectral transition between two one-sided band structures. A spectral flow or a change of topology requires additional interface gluing or a specified continuous regularization, as discussed in *Appendix C*.

### D. Non-ideal discontinuity and finite-order nonlinear

The preceding theoretical model describes DPMMs with ideal discontinuities. However, practical systems are subject to manufacturing tolerance eigenvalues within each partitions and material deformations, making perfect discontinuity unattainable, and the contact force at the atomic scale (repulsive force between electrons) varies continuously with the contact gap. Here, we introduce the transition length $\xi$ and the average transition rate $\chi$ to quantify the deviation of real DPMMs from the ideal model. They provide metrics for interpreting experimental results and demonstrate that DPMMs are highly nonlinear, if not perfectly discontinuous.

$\xi$ and $\chi$ *are* two dimensionless quantities that are applicable to compare an ideal stress-strain relation $\sigma(\epsilon)$ with first-order DS and an actual (infinite-order smooth) $\sigma_{ac}(\epsilon)$. If the linearity of $\sigma(\epsilon)$ is good at the both sides of the DS, whose slope are $\epsilon_1' = \lim_{\epsilon \to \mathrm{DS}_+} \sigma(\epsilon), \epsilon_2' = \lim_{\epsilon \to \mathrm{DS}_-} \sigma(\epsilon)$, respectively, then the transition length is the most direct criterion for evaluating the performance of DPMM. It is defined as:

$$\xi = \max\left(\left\{\epsilon_2 - \epsilon_1 \middle| \left|\frac{\sigma_{ac}(\epsilon_i) - \sigma(\epsilon_i)}{\sigma(\epsilon_i)}\right| > \mathcal{E}, i = 1{,}2\right\}\right) \tag{18}$$

where $\mathcal{E}$ is a chosen tolerance (here set to 0.01) that defines the acceptable deviation from the ideal bilinear response. This provides a practical metric to quantify the sharpness of the experimental stress-strain curves. However, if $\xi$ is too large or the linearity of $\sigma(\epsilon)$ is bad near the DS, average transition rate is a more general criterion for these cases. We suppose the actual stress-strain relation is a second-order smooth function $\sigma(\epsilon)$. $\chi$ is defined as the average of $\sigma''(\epsilon)$ from a position $\epsilon_0$ before contact to a position $\epsilon_0 + \Delta\epsilon$ after contact, compared to the smaller stiffness at the two positions $\sigma_{basic}' = \min\{\sigma_{ac}'(\epsilon_0 + \Delta\epsilon), \sigma_{ac}'(\epsilon_0)\}$:

$$\chi = \frac{1}{\Delta\epsilon\, \sigma_{basic}'} \left| \int_{\epsilon_0}^{\epsilon_0 + \Delta\epsilon} \sigma_{ac}''(\epsilon)\, d\epsilon \right| = \frac{1}{\Delta\epsilon} \frac{|\sigma_{ac}'(\epsilon_0 + \Delta\epsilon) - \sigma_{ac}'(\epsilon_0)|}{\sigma_{basic}'} \tag{19}$$

It indicates the "curvature" of the experimental stress-strain curves.

These quantities are later used to characterize axial stress-strain relation experimentally measured for several DPMM samples.

## III. Numerical verification

### A. Static analysis and stress-strain relation

1.Modeling method

We first analyze the static mechanical properties to verify the sharp change in tangent stiffness and Poisson's ratio near the contact transition. Considering a section of the DPMM sample shown in Fig. 2(a), the unit cell has the shape of a hollow arrow whose wings expand by pulling or retract by compressing. At rest, two parallel walls (blue) are placed on both sides of the array without contacting the wings. The walls limit lateral expansion.

We simulate 8 cells to verify the translational symmetry of the displacement field. In this model, a static unidirectional tensile displacement $|\boldsymbol{u}_{x,r\in S_{in}}\rangle = u_x$ is applied on the surface $S_{in}$ located at a cross-section of the DPMM, and the opposite surface $S_{out}$ is fixed (see Fig. 2(a)). In the initial situation with $u_x = 0$, the wings of the arrow just touch the side walls without stress. Then, if $u_x > 0$ a negative strain is generated, which expands the wings, and conversely $u_x < 0$ retracts the wings. The equivalent axial strain of the DPMM is defined as $\epsilon = \frac{u_x}{Na}$, where $a = 1.2$ cm is the cell length, and $N = 8$ is the number of cells in this simulated model. In our case, the DS is located at $\epsilon = 0$. The equivalent stress $\sigma$ on the DPMM is given by the axial force along the $x$-axis applied on the surface $S_{out}$ such that $\sigma = \int_{S_{out}} dS\, \sigma_{xx}/S_{out}$. The Young's modulus is then obtained with the slope of $\sigma(\epsilon)$. The vertical directional strain is obtained by extracting the displacement of the tip of the wing, which determines Poisson's ratio. More numerical details are written in *Appendix E*.

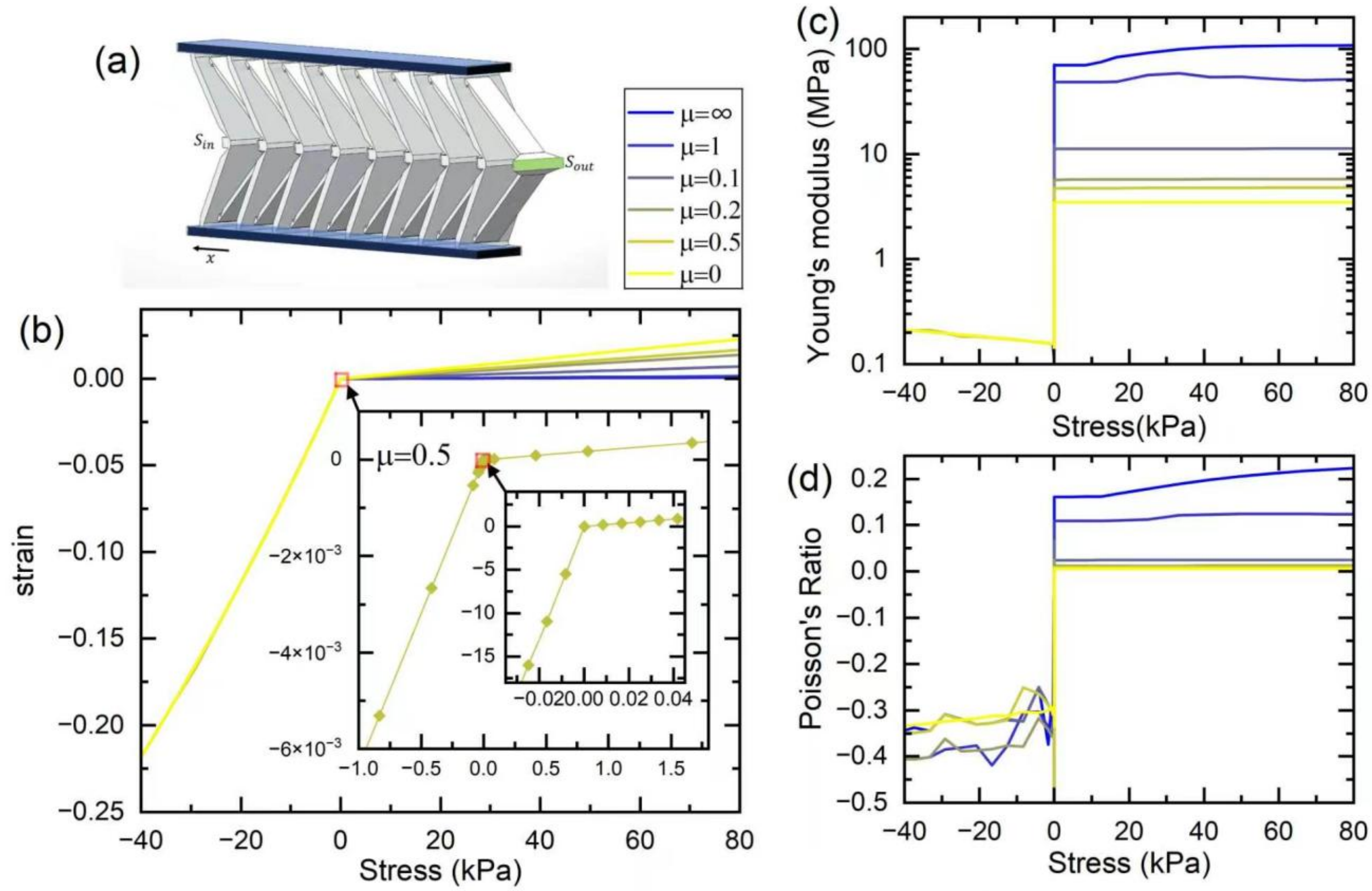

**Figure 2.** (a) Simulated arrow-type DPMM. (b) stress-strain relation with two zoom-ins showing the sharp inflection point at the DS. (c) Axial Young's modulus and (d) Poisson's ratio showing a discontinuity at the DS.

## 2. Results

Fig. 2(b) shows that the simulated $\sigma(\epsilon)$ is a continuous function, but non-smooth at $\epsilon = 0$, where the DS generates an inflection point. To validate the inflection is really non-smooth, two zooms were further performed around the DS. These zooms show that the inflection point always remains sharp, without a smooth transition. This indicates that under the simulation conditions, the displacement field can maintain translational symmetry, and the contacts of different cells occur simultaneously. Different friction coefficients between the walls and the wings were tested, ranging from frictionless with $\mu = 0$ to $\mu = \infty$, where the wings are stuck. When negative strain is applied, the friction has no effect since the wings retract. For positive strain, small differences appear as the friction changes the way wings slide against the wall, with the slope changes more as the friction increases. The non-smooth stress-strain relation must yield a discontinuity of the local Young's modulus $d\sigma/d\epsilon$ (This DPMM example is one-dimensional, the shear modulus is meaningless), as shown in Fig. 2(c). The different slopes in the positive part of the stress-strain relation clearly separate the Young's modulus values from around 4 MPa for $\mu = 0$ to 100 MPa for $\mu = \infty$. Fig. 2(d) shows that the DS is also a discontinuous point for the Poisson's ratio, since it is defined by directional strain. Poisson's ratio is negative when the wings retract under negative strain, and jumps to positive values when the wings contact the walls.

## B. Modal analysis and phonon band structure

Static analysis verifies the discontinuous stiffness and Poisson's ratio, which can lead to a discontinuity in the band structure. By considering the linear band structures of the two adjacent states before and after contact, we can intuitively see its discontinuous change. Through these adjacent spectra, we can intuitively see how easily the contact observability condition required for the eigenvalue jump in Sec. II.C is satisfied.

### 1. Modeling method

In this study, the following boundary conditions are adopted as complementary conditions to investigate the modes before the DS (non-contact case) and after (contact case), without the need to build an additional complementary eigenvalue problem solver. Here, the computational domain is the unit cell represented in Fig. 3(a), which is also the DPMM investigated in the previous section. Bloch periodic conditions are applied along the x-direction to simulate an infinitely long one-dimensional system. Then, we use $\boldsymbol{k} = k_x \in [0, \pi/a]$ sweeping half of the first Brillouin zone simplified by the inversion symmetry. The non-contact and contact cases are studied by defining different conditions at the extremities along the y-direction of the unit cell, which represents how the surface-to-surface contact of DPMM interact with the wall. Equivalent elastic boundary conditions are applied at the contact interface between the wing and the wall, with normal and tangential stiffnesses denoted as $E_n$ and $E_t$, respectively, to simulate the constraint of the contact on the unit cell.

Three cases were studied. First, the non-contact case, where the wings are unconstrained with $E_n = E_t = 0$, equivalent to free boundary conditions, considering that the walls on either side are removed. Then, two contact cases are studied. A frictionless case where the wings can slide freely along the x-direction, but they cannot expand. In this case, $E_n = 100$ GPa and $E_t = 0$. Last, a rough contact case where the extremities of the wings are fixed using $E_n = E_t = 100$ GPa. The value 100 GPa is on the order of the Young's modulus of common alloys, simulating the elastic deformation of the material after contact. It should be noted that the above is not a simple parameter comparison; from the non-contact case to the contact case, only an infinitesimal strain needs to be applied. We can investigate the eigenvalue jump like Eq. (13) described through this comparison.

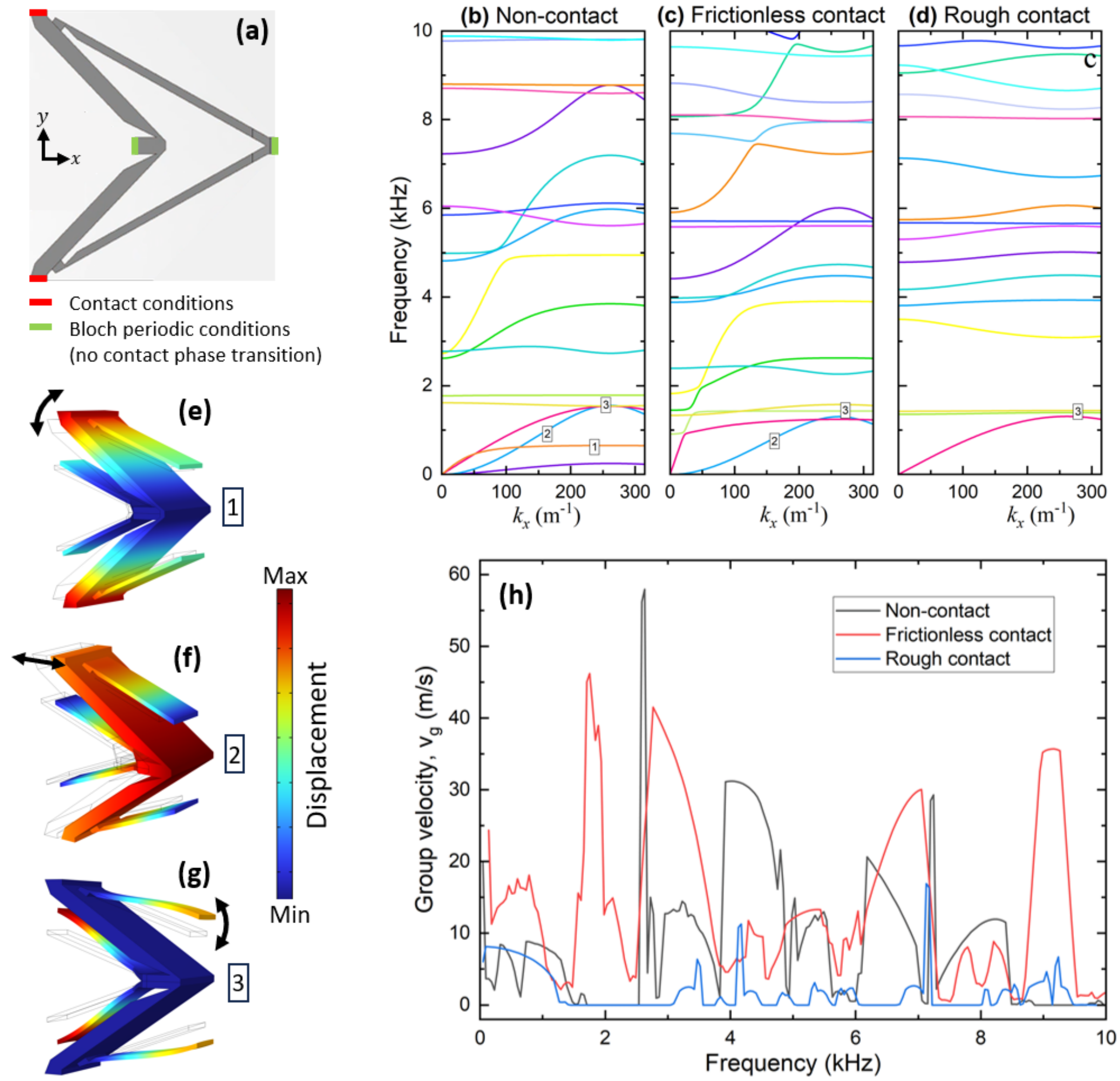

**Figure 3.** (a) DPMM unit cell. Phonon band diagram for the non-contact case (b), the frictionless contact case (c), and the rough contact case (d). (e), (f) and (g) Examples of low-frequency modes. (h) Spectral phonon group velocity in the three cases.

## 2. Results

Fig. 3(b) to 3(d) are the eigenfrequency spectrums of the above 3 cases at frequencies up to 10 kHz. Adding constraints at the extremities limits the possible displacement leading to forbidden displacement modes. This is clearly visible in low frequency and especially on acoustic bands where 4 bands are observed in Fig. 3(b), then 2 bands in Fig. 3(c), and then one in Fig.3(d). To understand this phenomenon, we observed the modal displacement field (eigenvectors) at selected $k_x$ locations across the cases. Table 1 summarizes the presence of modes in the different cases. Mode 1 is a wing-expansion mode where wings can swing externally. Mode 2 is a x-translation mode where the whole cell can oscillate along the x-axis. Mode 3 represents the plate distortion. Mode 3 is present in all cases as it is independent of the contact conditions. Mode 2 appears in non-contact and frictionless-contact cases but not in rough-contact cases because the wings cannot move, whereas mode 1 can only exist in the non-contact case and is forbidden in other cases due to the walls. Apart from the generation and disappearance of some modes, finite jumps in modes are more common. Such as the translation mode 2 changes its eigenfrequency from non-contact to frictionless contact. It is because at finite $k_x$, mode 2 may no longer be a pure rigid-body translation, but may mix in wing rotation, bending or normal deformation, which can observe the contact and makes $\mathfrak{C}_{\mathrm{act}}(|\boldsymbol{u}\rangle, |\boldsymbol{u}\rangle) > 0$ under the description scope of Eq. (12). In conventional topological phase transition systems, although the topological properties (such as the Chern number) of the eigenvector bundle can undergo discontinuous changes in some symmetry protected cases, the eigenvalue spectrum is still continuous on the parameter space.[23,53,61,62] By contrast, DS are not only singularities of the eigenvector bundle, but also singularities of the eigenvalue spectrum.

**Table 1:** Presence of modes for different cases.

| Mode \ Case | Non-contact | Frictionless contact | Rough contact |
|---|---|---|---|
| Wing expansion | ✓ | X | X |
| x-translation | ✓ | ✓ | X |
| Plates distortion | ✓ | ✓ | ✓ |

The discontinuous property change in the band structure also reflects on the transport properties via the average group velocity, following:

$$v_g(\omega) = \frac{1}{\pi\, D(\omega)} \sum_m \int_0^{\pi/a} |\partial_{k_x}\omega_m|\, \delta\big(\omega - \omega_m(k_x)\big)\, dk_x \tag{20}$$

where $D(\omega) = \frac{1}{\pi}\sum_m \int_0^{\pi/a} \delta\big(\omega - \omega_m(k_x)\big)\, \mathrm{d}k_x$ is the density of states, which also changes discontinuously (see *Appendix G*). Fig. 3(h) shows that $v_g$ is generally higher in the non-contact case than in the frictionless contact case, since the bands are flatter and behave “less softly” when a cell is not contacting the wall. Without quantifying the specific energy transfer distribution caused by phonon scattering, these overall transport properties derived from the phonon spectra on both sides of the contact can roughly predict the behavior of mechanical waves across the DS.

It should be noted that Table 1 is intended to demonstrate how contact constraints qualitatively alter the structure of the modal space, and the mode classification therein is obtained under the assumption of uniform contact. In a real DPMM, the contact phase change may be frequent, and the excitation and propagation of modes will be more complex. However, at a DS the tangent operator is not defined by a single smooth linearization; uniformly periodic contact can still retain Bloch periodicity, whereas a nonuniform contact pattern or a transient crossing requires a nonlinear transient calculation.

### C. **Dynamics near discontinuity**

Modal analysis captures the linear wave characteristics on either side of the DS but fails at the DS itself due to the system's nonlinear behavior there. To investigate the dynamic behavior of DPMMs, which are the vibrations across the DS, the full nonlinear transient dynamic equations must be solved. This approach reveals strong nonlinear effects driven by discontinuity, particularly strong intermodal coupling.

#### 1. Modeling method

The wave transport properties of ordinary materials depend on the Young's modulus $E = \frac{\kappa(3\lambda+2\kappa)}{\lambda+\kappa}$ and the density $\rho$. The speed of sound in isotropic solids is given by $\sqrt{E/\rho}$, but since DPMMs properties change across with contact, sudden change in sound speeds is expected. To investigate this effect, the speed of sound in a DPMM is simulated with the propagation of a pulse through the DPMM. Furthermore, phenomena of wave mixing are also studied with the response under multi-frequency harmonic strain input to observe intermodal coupling.

Here, a similar DPMM design as shown in Fig. 2(a) is used but with number of cells $N = 100$ and $\mu = 0.3$. A time-dependent strain pulse, as shown in the inset of Fig. 4 (a), is applied at the input section $S_{in}$ in the axial direction $\epsilon_x$. The output section $S_{out}$ is fixed, as previously. At $t = 0$, the input perturbation starts with $\epsilon = \epsilon_{x0}$, and moves in axial direction with constant strain rate $\frac{0.01\ [\text{m/s}]}{Na}$, eventually stops at $\epsilon = \epsilon_{x0} + \frac{1[\mu\text{m}]}{Na} = \epsilon_{x0} + 8.33 \times 10^{-7}$ at $t = 0.1\ ms$. The strain $8.33 \times 10^{-7}$ is chosen to probe the theoretical sensitivity of the model. While such precision is challenging to achieve experimentally, it serves as a numerical benchmark to illustrate the potential for strong response at small perturbations. The time $t_0$ for a specific mode to reach the opposite end can be obtained by detecting the specific signal at $S_{out}$, as detailed below. Then, the speed of the quasi-mode is derived by $Na/t_0 = L/t_0$, where $L = 1.2$ m is the total length. The accuracy of this approach is verified in *Appendix F* with a simple solid rod.

Multi-frequency harmonic strain wave propagation is simulated to verify the strong nonlinearity near the DS. The simulation model is identical to the pulse, but the boundary conditions at the input and output are replaced by time-dependent harmonic functions. At input section $S_{in}$, we use the displacement $|\boldsymbol{u}_{x,r\in S_{in}}(t)\rangle = A_1\sin(2\pi f_1 t) + A_2\sin(2\pi f_2 t) + u_0$, $|\boldsymbol{u}_{y,r\in S_{in}}\rangle = |\boldsymbol{u}_{z,r\in S_1}\rangle = 0$to simulate the interaction of two harmonic excitation. where $A_1 = 1$ μm, $u_0 = 8$ μm and $f_1 = 53\ Hz,\ \ f_2 = 351\ Hz$ are fixed.

Subsequently, only the parameter $A_2$ is changed between 0 and 0.3 μm. The input strain variation is kept at $8.33 \cdot 10^{-7}$, like previously. At $S_{out}$, a damping $\partial_t |\boldsymbol{u}_{x,\boldsymbol{r}\in S_{out}}\rangle = 1.84\,\mathrm{N/m \cdot s}$ is applied to reduce the reflectivity of the boundary, especially the reflectivity of the x-direction translation mode. Take the section 20 cells away from $S_{in}$ as $S_{out2}$, the average displacement of output section $\langle |\boldsymbol{u}_{x,\boldsymbol{r}\in S_{out2}}|\rangle$ is extracted as the output signal. The equivalent strain $\epsilon$ is directly proportional to $\langle |\boldsymbol{u}_{x,\boldsymbol{r}\in S_{out2}}|\rangle$ . The output strain is then plotted as a function of time up to 0.3 s for the two cases: (i) non-contact, (ii) $\mu = 0.3$. Frictionless contact is not used here because of the numerical difficulty. Finally, a fast Fourier transform (FFT) of each signal in the steady state is performed to analyze the contribution of each vibration (phonon) modes.

## 2. Results

The propagation speed of the x-translation mode and the wing swinging mode are calculated. The x-translation mode is detected by the stress mutation at $S_{out}$. The wing swinging mode is detected by the contact gap of the last cell, where the emergence of a gap means the phase transition from contact to non-contact. Fig. 4(a) shows that the speed of the x-translation mode is rather stable at 370 m/s because the contact does not directly prevent the x-translation movement. The maximum group velocity of the lowest 30 bands of an x-translation mode of no-contact, as calculated in Fig. 3(h), is marked in dashed lines in Fig. 4(a). The contact behavior makes the signal speed significantly faster since the contact makes the cells stiffer and establishes interactions among many other modes with higher group velocity. On the other hand, the speed of wing swinging mode shows high sensitivity to strain and exhibits wide variations due to the random-like contact behavior among the many cells due to the nonlinearity of the DPMM. This is illustrated in Fig. 4(b) with a snapshot of the pulse propagation through the DPMM. The cells after the core of the signal are randomly deformed with mixed states of contact and non-contact. It yields a high complexity in the mechanical wave components and rapid non-equilibrium phase transition.

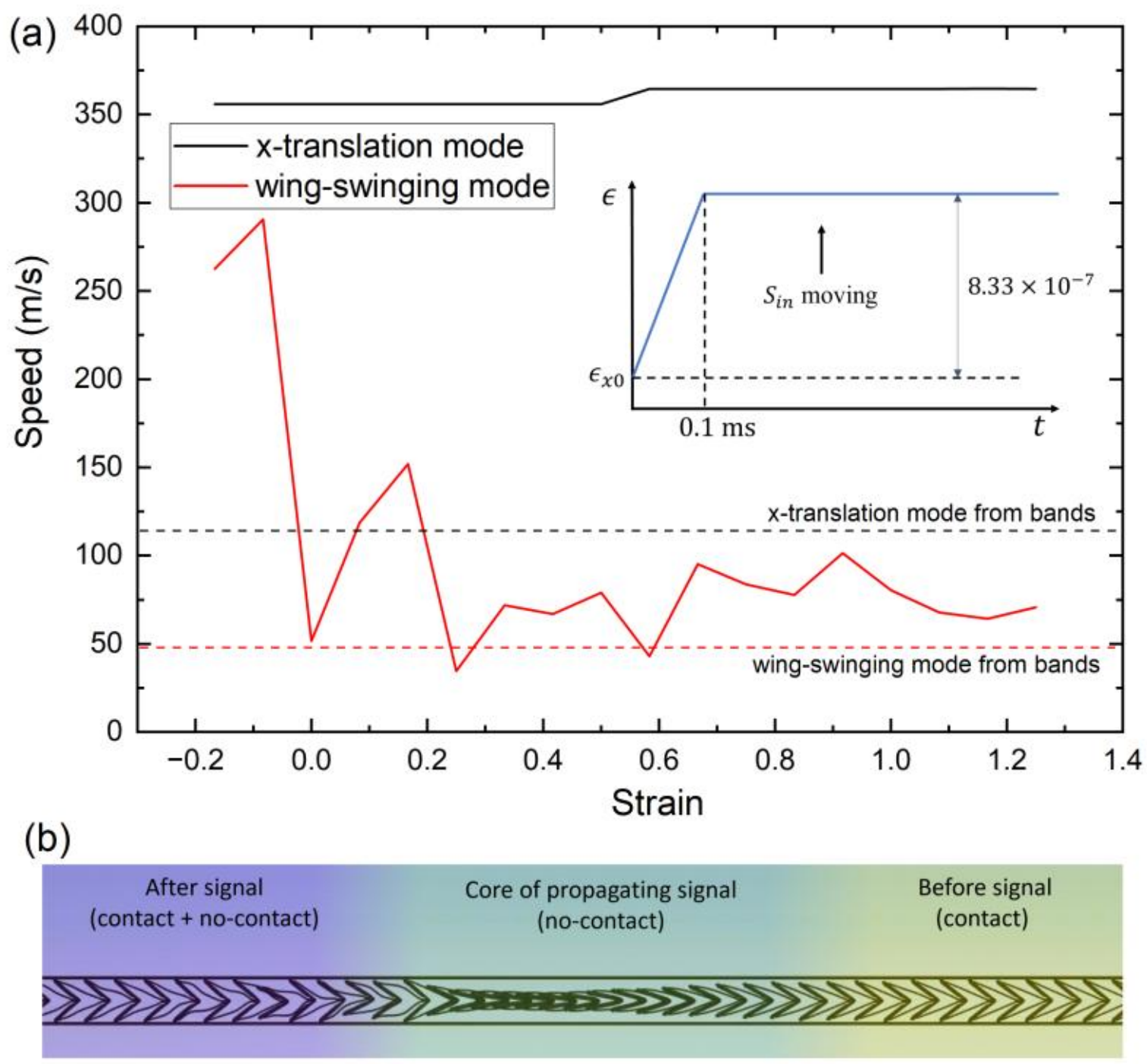

**Figure 4.** (a) Propagation speed of x-translation mode and wing swinging mode when the DPMM is submitted to a pulse like shown in the inset. (b) Snapshot of the pulse propagation in the DPMM showing the cell deformation and the contact state before, during, and after the passage of the signal.

In Fig. 5, excitation with multi-frequency harmonics is used to investigate the complex wave mixing components behavior. Using the non-contact case as the comparison, Fig 5(a) shows that the output strain follows the input harmonic excitation, which is confirmed by the fast Fourier transformation (FFT) in Fig. 5(b) where only the frequencies $f_1$ and $f_2$ appear. When the side walls are included with a friction $\mu = 0.3$, the output strain still shows periodic-like pattern, as shown in Fig. 5(c). However, the FFT in Fig. 5(d) shows no presence of the input frequencies. Instead, some new frequency peaks like $f_c = 35\ Hz$ appears, which corresponds to no combination of $f_1$ nor $f_2$. Higher harmonics of $f_c$ appear with the intensity of these harmonics being amplified and the intensity of $A_2$ increases and their frequency position shifts to lower frequencies as the harmonics get higher. Calculate $\Upsilon = \int_{f\in[400\text{Hz},8333\text{Hz}]} \text{FFT}(\epsilon)\, df$ to estimate the intensity of phonon scattering at higher frequency range. The value is marked in Fig. 5(b), (d). For the non-contact case, this value is theoretically equal to zero. Its magnitude indicates the numerical noise. In Fig. 5(c), when $\mu = 0.3$ contact is introduced, all $A_2$ cases give $\Upsilon$ an order of magnitude larger, clearly distinguishable from numerical noise. The emergence of new frequency components $f_c$ and their harmonics demonstrates strong nonlinear wave mixing in the DPMM. This can be interpreted as efficient mode conversion and energy redistribution among vibration modes, analogous to phonon scattering in anharmonic lattices, but here driven directly by the contact nonlinearity.

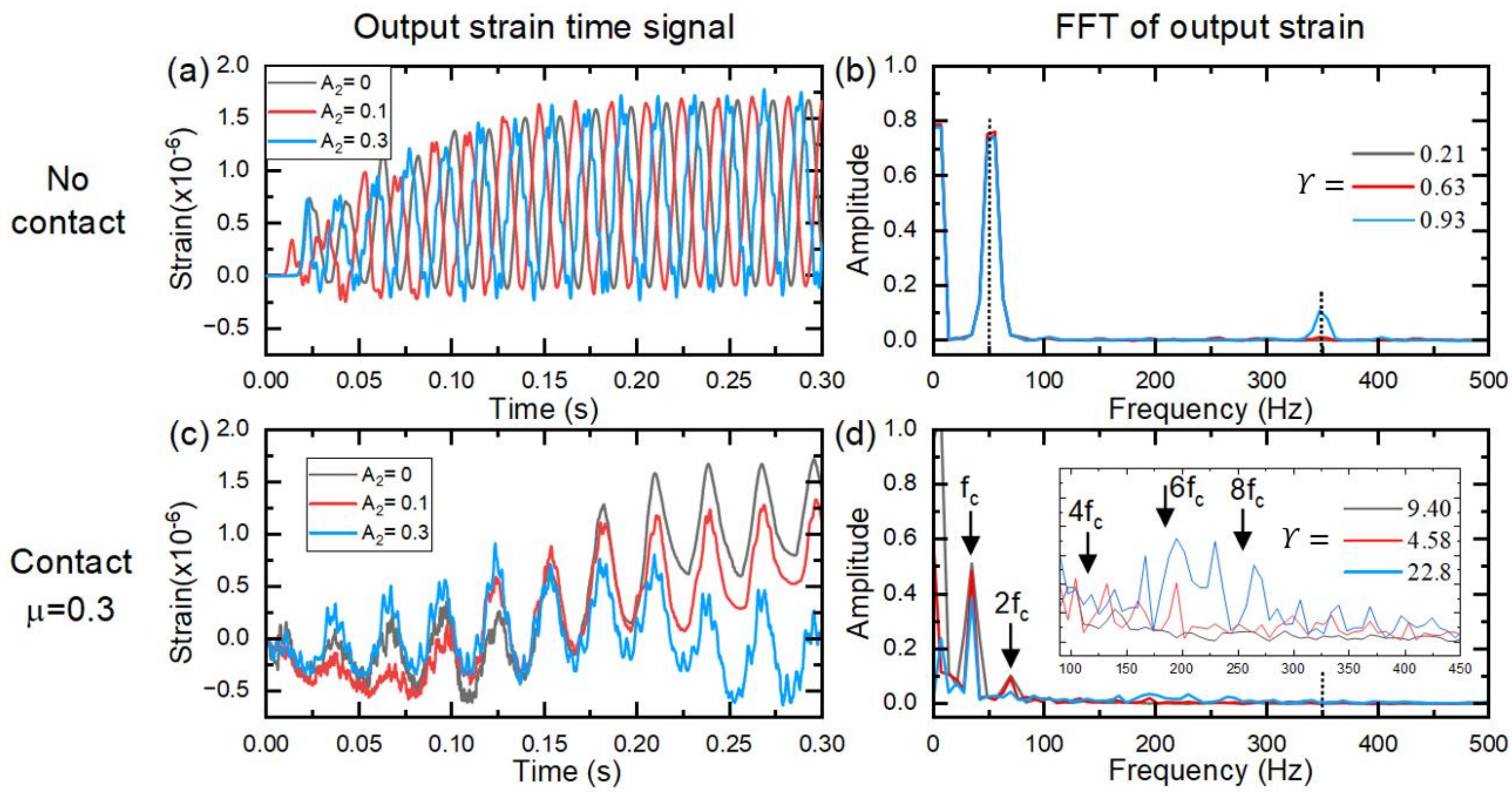

**Figure 5**. Time domain output signal of (a) non-contact and (c) $\mu = 0.3$ contact. Fourier transformation of the output signal (relaxation section removed) with (b) non-contact and (d) $\mu = 0.3$ contact.

The results of the two dynamic examples above demonstrate that DPMM exhibits strong nonlinearity through classical mode coupling, equivalently interpretable as many-phonon interaction after quantization. When a phonon propagates through a DS, its own amplitude can trigger a phase transition between non-contact and contact states. The stiffness jump causes the lattice background on which the phonon depends to change drastically and instantaneously during propagation. The abrupt change in the background forces the system to redistribute the phonons' energy and momentum, leading to strong inelastic scattering, in which phonons split, merge, and generate new frequency components [63]. Just as a periodic mechanical structure with a second-order contact model can exhibit the second harmonic[64]. The ideal (discontinuous) contact model may generate infinitely many nonzero harmonics; their spectral decay depends on the time regularity and the contact law. In conventional materials, phonon scattering caused by defects or inherent lattice vibrations, which disrupt periodicity, is also common[65,66]. However, the contact nonlinearity in DPMMs is so strong that even a strain perturbation of the order of $10^{-5}$ can cause different phonon scattering behaviors. On the other hand, phonon scattering can be regarded as the cause of the variation in properties of DPMM, just like the speed of light in the medium is affected by photon-electron coupling.[67]

An ubiquitous phenomenon caused by nonlinearity is chaos, and relevant studies have been conducted in both acoustic[68] and optical[69] systems. Because the strain of DPMM is a nonlinear parameter, the speed of the wing-swing mode sensitively depends on the strain (Fig. 4(a)) and induces randomness. In fact, all the conditions and simulation processes are deterministic. In the DPMM, each minute initial-strain adjustment can transition the system between distinct metastable contact configurations, such as switching from contact to separation or from sliding to sticking. Thus, macroscopically smooth strain variations guide the system along a highly path-dependent route into a different local energy minimum, each characterized by unique effective stiffness and force distribution.

## IV. Engineering Feasibility

In reality, various factors will inevitably destroy the discontinuity. This extended section intends to test whether the discontinuity-like behavior of DPMMs is easy to maintain in practice. Now, we test the strain-stress relation of a simple DPMM example experimentally to observe the impact of the non-ideal geometry and dynamics. As an additional section, the experiments are not intended to validate the theoretical framework or the numerical simulations presented above, instead, they serve as a separate proof-of-concept demonstration of the practical feasibility of constructing a DPMM with the designed nonlinear response.

### A. Experimental Methods

DPMMs of the external-contact type, as studied in previous sections, were fabricated in resin via 3D printing. A custom-made metal frame to support the DPMM and limit the external expansion of the wings was fabricated and equipped with a micrometer screw gauge to apply strain and a pressure sensor to measure the stress. The range of the micrometer screw gauge allowed for pulling up to $-75$ μm and to push up to $475$ μm, with a resolution of $25$ μm, corresponding to a strain between -0.0014 and 0.0090 and a resolution of $4.73 \times 10^{-4}$. The length of the frame supported a DPMM with 11 cells. The aim is to experimentally obtain the stress-strain relation, to observe the discontinuity-like stiffness and verify that the DPMM can achieve a high degree of nonlinearity in actual scenarios.

We built platform with a force sensor to measure the strain-stress relationship as shown in Fig. 6(a). DPMM samples were fabricated using a 3D printer FormLabs Form 3 with a grey v4 resin. A rough cleaning process was used. A translation block with accuracy 8.33 μm is used to tune the strain. One end of the sample is fixed, and at the other end, a 5 N range force sensor is installed on the sample's connector. There is no contact between the sample and the translation block on the force sensor side. On the other side of the sample's connector, it directly contacts the translation block, which can be regarded as a stiff spring and is much stiffer than the sample itself, so the deformation and the change in pushing force of the spring can be ignored when applying a small strain on the sample. This ensures that the sensor's force variation matches that of the sample.

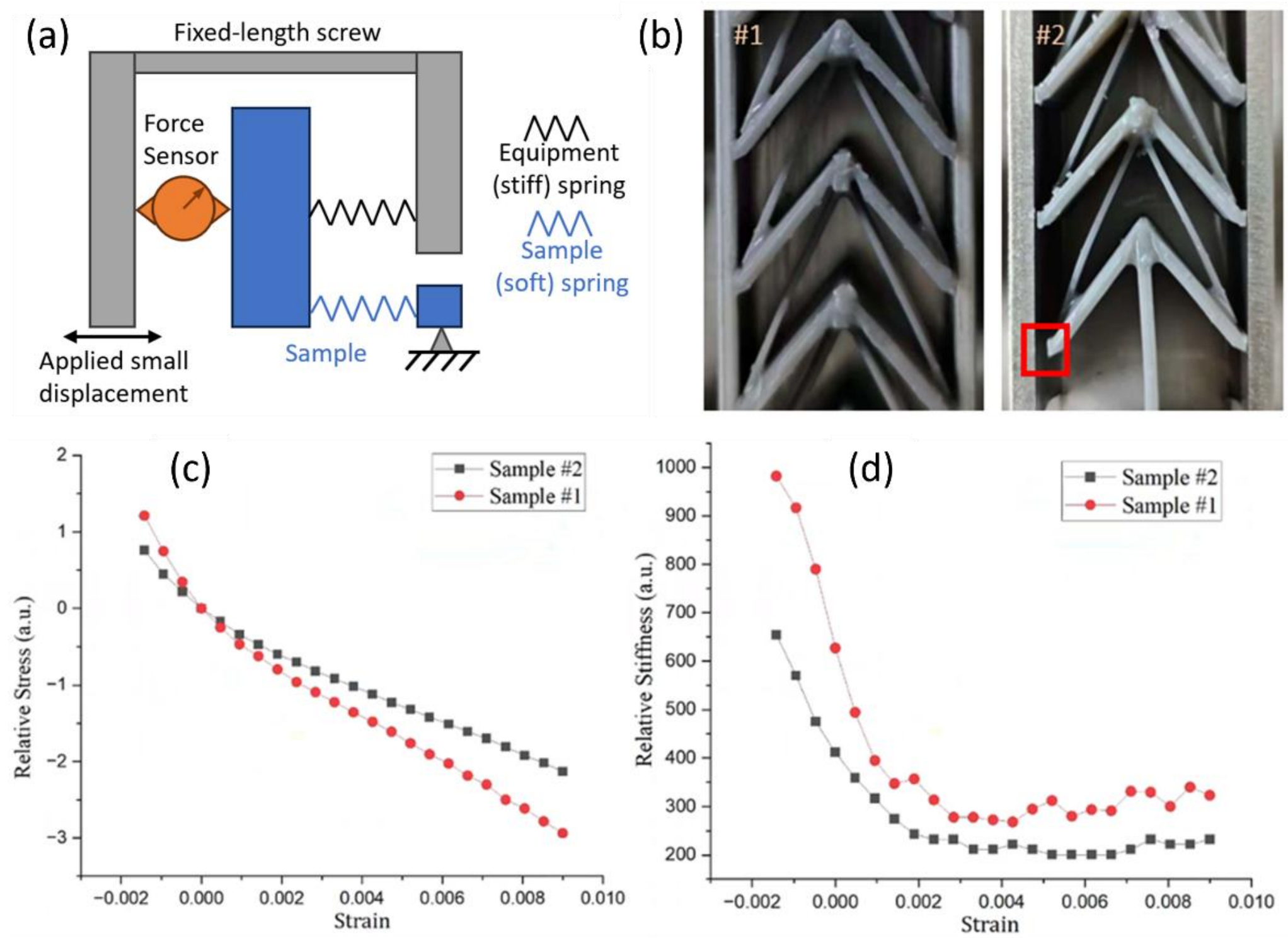

**Figure 6.** (a) Schematic of the experimental setup. (b) 3D printed DPMM examples where the sample #1 is good and the sample #2 has defects (see red square). (c) Measured stress-strain relation. (d) Measured Stiffness.

### B. Experimental results

Two samples (#1 and #2) were fabricated with the same theoretical design but different fabrication configurations. Sample #1 has a better quality than the sample #2. No clear correlation was established between the printing parameters and the quality. An example of defect with a wing smaller than the others in the sample #2 is shown in Fig. 6(b). Figs. 6(c) and 6(d) show the experimental stress-strain relation and its strain derivative (stiffness), respectively. Starting from a strain of -0.0014 (limited by the device's pulling force), we slowly increase the strain from negative (pulling) to positive (pushing). In this process, the 11 cells almost deform simultaneously, making it difficult to discern the sequence of the different cells' contacts with the eye. This shows that, due to finite fabrication precision and non-simultaneous contact, the ideal discontinuity is regularized into a sharp but continuous transition over a small strain interval. The measured transition length ξ and average transition rate χ quantify this regularization, confirming that the DPMM concept remains highly nonlinear, with a stiffness variation by a factor of three over a strain range of $10^{-3}$.

To quantitatively evaluate the experimental DPMM performance, the transition length ξ and average transition rate $\chi$ defined previously were used. The sample #1 obtained $\xi \approx 0.003$, and $\chi \approx 4.0$ while the sample #2 obtained $\xi \approx 0.004$, and $\chi \approx 3.5$. It means that for samples with more than 13 cm length, the stiffness can increase by more than three times under a total deformation of less than 400 micrometers. The stiffness difference between samples #1 and #2 may be due to manufacturing process errors, such as photopolymerization. Despite the defects in sample #2 the discontinuous-like behavior was still clear. This demonstrates the high nonlinearity of given DPMM example is accessible experimentally even with low-cost fabrication and that their performance is quite robust to the conditions of fabrication.

## V. Conclusion and Remarks

This work introduced the concept of Discontinuous Property Mechanical Metamaterials (DPMMs), a class of macroscopic solid structures that exploit designed mechanical discontinuities to achieve extreme nonlinearity. We established a theoretical mechanical framework suitable for contact discontinuity conditions. When the non-contact equilibrium crosses zero gap transversely, the parameter discontinuity set is a smooth hypersurface of the finite-dimensional parameter space. For finite activated stiffness, contact-observable simple modes have eigenvalue and eigenvector jumps under the stated condition. In order to verify our theoretical predictions, we validate discontinuous stiffness and Poisson's ratio, observe phonon dispersion, and enabling strong nonlinear wave mixing and mode conversion with numerical simulation. A simple experiment confirms that this highly nonlinear property is achievable in practical systems, even under simple manufacturing conditions. Our research establishes a foundational research framework for engineering mechanical discontinuities in metamaterials, bridging macroscopic mechanical contact with nonlinear dynamic of interacting system. This opens new avenues for applications that require extreme nonlinear dynamic response.

**Acknowledgments**

This work was supported by the Shenzhen Science and Technology Program (No. JCYJ20240813141502004).

**Data availability**

The data that support the findings of this study are available from the corresponding author upon reasonable request.

## *Appendix A: Introduction to dynamical system with discontinuous stiffness*

The dynamics of non-smooth potentials have been well studied using established theories. As is well known, nonlinearity is often accompanied by phase-trajectory bifurcations and chaotic phenomena. As the limiting case of nonlinearity, discontinuity naturally gives rise to chaos. Compared with the "finite-degree" nonlinearity in smooth systems, it exhibits several distinct dynamical characteristics.

The first characteristic is that a stable periodic orbit can directly transition to chaos via collision with a discontinuous boundary, without undergoing a period-doubling cascade. Nusse et al. [36] first systematically defined border-collision bifurcations (BCB) and provided an example of a one-dimensional piecewise linear potential. When a parameter was tuned to a certain threshold, chaos emerged abruptly, without the gradual progression from period-two and period-three orbits to chaos as observed in smooth nonlinear systems. Recent studies [39,70] further support this conclusion. Engineering examples such as power electronic converters in Ref. [71] demonstrate in detail how BCB directly induces chaos. Ref. [38] comprehensively describes the bifurcation theory of discontinuous systems and clearly states that BCB serves as a shortcut to chaos, eliminating the need for period-doubling cascades. This occurs because, when a stable periodic orbit intersects a discontinuous boundary transversely, its Poincaré return map exhibits a non-smooth, piecewise hyperbolic structure. Under transversality and unfolding conditions, this map can directly generate topological horseshoes, thereby producing chaotic invariant sets. Since discontinuity allows multipliers to jump instantaneously outside the unit circle, the system can enter chaos directly without a cascade of period-doubling bifurcations.

The second characteristic is the frequent coexistence of multiple attractors near bifurcation points. Dutta et al. [37] proposed the concept of multi-attractor bifurcation and proved that multiple stable attractors can arise simultaneously in BCB. The abrupt change in the vector field caused by phase trajectories crossing the boundary tears the basins of attraction into fractal structures. Furthermore, Simpson et al. [35] proved theoretically that multiple disjoint attractors can exist during BCB in phase space at the same parameter value, and that such configurations can bifurcate directly from a stable fixed point. In this case, the chaotic attractor is no longer a continuous manifold in phase space as in smooth systems, but appears as fragmented patches. Initial conditions leading to multiple attractors coexist within any arbitrarily small neighborhood, with their boundaries fractured and interwoven[72]. This phenomenon can be observed in gear transmission systems[73]. One consequence of multi-attractor coexistence is the robustness of chaos: chaotic attractors reside within open sets in parameter space, unlike chaos in typical smooth nonlinear systems, whose chaotic parameter intervals contain periodic windows. Based on this phenomenon, Banerjee et al. [74] put forward the concept of robust chaos. Ghosh et al. [75] extended the proof of the existence of chaos to non-invertible and orientation-reversing piecewise linear maps, revealing that robust chaos is independent of global topological properties.

### *Appendix B: Ellipticity of the operator $\mathcal{L}$*

In definition, $\langle \mathcal{L}|\boldsymbol{u}\rangle|\boldsymbol{v}\rangle = a\,(|\boldsymbol{u}\rangle, |\boldsymbol{v}\rangle) = \int_\Omega \widehat{\boldsymbol{\sigma}} : \hat{\boldsymbol{\epsilon}}\ d\boldsymbol{r},\ \forall|\boldsymbol{v}\rangle \in \mathcal{K}_{\text{nc}}$. In elastic solid Eq. (3), the principal part of $\mathcal{L}$ is

$$\mathcal{L}_{\text{prin}} = \kappa\boldsymbol{\nabla}^2 + (\kappa+\lambda)\boldsymbol{\nabla}(\boldsymbol{\nabla}\,\cdot) \qquad (B1)$$

Taking the Fourier transform, the symbol matrix $\boldsymbol{L}(\boldsymbol{\xi})$ is:

$$\boldsymbol{L}\,(\boldsymbol{\xi}) = \kappa|\boldsymbol{\xi}|^2 I + (\kappa+\lambda)\,\boldsymbol{\xi}\boldsymbol{\xi}^T$$

For any non-zero $\boldsymbol{\xi} \in \mathbb{R}^3$ and any non-zero vector $\boldsymbol{w} \in \mathbb{C}^3$,

$$\boldsymbol{w}^*\,\boldsymbol{L}\,(\boldsymbol{\xi})\,\boldsymbol{w} = \kappa|\boldsymbol{\xi}|^2|\boldsymbol{w}|^2 + (\kappa+\lambda)|\boldsymbol{\xi}\cdot\boldsymbol{w}|^2$$

By stability conditions $\kappa > 0$ and $\kappa + \lambda > 0$, this quadratic form is strictly positive. Hence the principal symbol is positive definite: there exists a constant $c > 0$ such that

$$\boldsymbol{w}^*\,\boldsymbol{L}\,(\boldsymbol{\xi})\,\boldsymbol{w} \geq c|\boldsymbol{\xi}|^2|\boldsymbol{w}|^2 \qquad (B2)$$

This is the definition of strong ellipticity for a second-order system. Therefore, $\mathcal{L}$ is an elliptic operator.

This conclusion is not limited to the isotropic case. For principal symbol for general Linear Elasticity $\boldsymbol{A}_{ik}\,(\boldsymbol{\xi}) = \boldsymbol{C}_{ijkl}\,\boldsymbol{\xi}_j\,\boldsymbol{\xi}_l$, strong ellipticity is defined as

$$\boldsymbol{w}^*\,\boldsymbol{A}\,(\boldsymbol{\xi})\,\boldsymbol{w} > 0, \forall\boldsymbol{\xi}, \boldsymbol{w} \neq 0 \qquad (B3)$$

Such as for incompressible hyperelastic models like the Neo-Hookean, Mooney-Rivlin, and Ogden models, the ellipticity condition may be satisfied or lost under specific deformation conditions[76].

### *Appendix C: Suggestion of Discontinuous Band Topology Studies*

As mentioned in Sec. II.C, the parameter space $\mathcal{P}$ is partitioned by the parameter switching set $\hat{\Sigma}$ into smooth connected regions $\mathcal{P}_{\text{j}}$, and the modal bundles $M_{\text{m}}$ are defined on $\mathcal{P} \setminus \hat{\Sigma}$. The contact problem is described by the complementary eigenvalue problem introduced in Eq. (4), while its weak formulation and ellipticity properties are given in Eqs. (10) and (B3).

The main purpose of this Appendix is to clarify how the modal bundles on the two sides of $\hat{\Sigma}$ can be related and under which conditions this relation carries topological information. The conditional simple-mode and modal-cluster jumps at $\hat{\Sigma}$, established in Sec. II.C, do not by themselves define a topological invariant. A topological characterization requires the study of the isolated modal subspaces and of their gluing across $\hat{\Sigma}$.

**C.1. Smooth modal bundles on the partitioned parameter space**

On each smooth region $\mathcal{P}_{\mathrm{j}}$, the contact configuration is fixed and the complementary eigenvalue problem reduces to a smoothly parameterized elliptic eigenvalue problem. The continuity and analyticity properties of isolated eigenvalues and spectral subspaces then follow from standard perturbation theory for self-adjoint operators [50,51]. Consequently, the modal bundles $M_{\mathrm{m}}$ are well defined on each $\mathcal{P}_{\mathrm{j}}$, provided that the corresponding eigenvalue or isolated group of eigenvalues remains separated from the rest of the spectrum.

For a non-degenerate band, the Berry connection and Berry curvature introduced in Sec. II.C are therefore defined on each smooth region $\mathcal{P}_{\mathrm{j}}$. The same construction applies to an isolated group of bands by replacing the individual eigenstate with the corresponding modal subspace. In this case, the Berry connection is generally non-Abelian, whereas its curvature and Chern character are defined from the associated finite-rank spectral bundle.

The bulk topological information can thus be assigned separately to each smooth region. For example, if a closed surface is contained entirely within $\mathcal{P}_{\mathrm{j}}$ and the selected modal group remains isolated on this surface, its first Chern number is obtained from

$$\mathcal{C} = \frac{1}{2\pi}\int_{S} Tr\Omega^{P} \qquad (C1)$$

The integral in Eq. (C1) is meaningful only when the modal subspace is globally defined on the integration surface. It must therefore be evaluated separately on the smooth regions and cannot be extended across $\hat{\Sigma}$ without specifying how the two sides are related. Here $S$ is a closed oriented surface on which the selected spectral subspace is globally isolated.

For the real elastic operator considered here, the relevant symmetry class must also be taken into account. In the absence of gyroscopic, rotational, or other time-reversal-breaking terms, the eigenproblem can often be represented by a real operator. In that case, ordinary complex Chern numbers may vanish or fail to distinguish the relevant phases. The corresponding topology may instead be described by real vector bundles and their associated $\mathbb{Z}_2$, Stiefel–Whitney, or real $K$-theory classes. The choice of invariant is therefore determined by the symmetry of the operator family rather than by the existence of a contact discontinuity alone. Accordingly, the $U(N)$ gluing construction below applies after complexification or when a complex Bloch representation is used; if a globally real gauge is imposed, the structure group is $O(N)$ ($O(1)$ for a single real band), and real-bundle invariants must be used instead [23,24].

**C.2. Modal overlap at the discontinuity set**

Consider two adjacent smooth regions separated by $\hat{\Sigma}$, and let $|\boldsymbol{u}_m^{\mathrm{c}}\rangle$ and $|\boldsymbol{u}_m^{\mathrm{nc}}\rangle$ denote the one-sided modal limits obtained from the contact and non-contact configurations, respectively. For a common isolated collection of modes, define the modal overlap matrix $O_{\hat{\Sigma}}$ by

$$[O_{\hat{\Sigma}}]_{mn} = b(|\boldsymbol{u}_m^{\mathrm{c}}\rangle, |\boldsymbol{u}_n^{\mathrm{nc}}\rangle), \;\; \boldsymbol{P} \in \hat{\Sigma} \tag{C2}$$

where $b$ is the mass bilinear form introduced in Eq. (10), extended to complex Bloch modes in the usual Hermitian manner.

The matrix $O_{\hat{\Sigma}}$ measures the geometric mismatch between the two modal subspaces. For individually tracked, non-degenerate modes in a fixed gauge, its diagonal elements quantify the persistence of individual modal channels, whereas its off-diagonal elements describe the mixing between modes that are separated on the two sides of the discontinuity. These individual matrix elements are otherwise basis dependent within a degenerate modal cluster. The singular values of $O_{\hat{\Sigma}}$ are independent of the particular orthonormal bases chosen within the two modal subspaces and provide a basis-independent measure of their relative orientation.

The overlap matrix should be distinguished from the stiffness-based modal transition matrix $R$. The latter describes the dynamical response produced by the contact-induced stiffness change, whereas $O_{\hat{\Sigma}}$ describes the geometric relation between the modal subspaces. Consequently, a non-diagonal $R$ demonstrates contact-induced modal conversion, but it does not by itself establish a non-trivial topology.

If $O_{\hat{\Sigma}}$ is square and invertible, its polar decomposition gives

$$O_{\hat{\Sigma}} = Q_{\hat{\Sigma}}\left(O_{\hat{\Sigma}}^{\dagger} O_{\hat{\Sigma}}\right)^{1/2} \tag{C3}$$

with

$$Q_{\hat{\Sigma}} = O_{\hat{\Sigma}}\left(O_{\hat{\Sigma}}^{\dagger} O_{\hat{\Sigma}}\right)^{-1/2} \tag{C4}$$

The factor $Q_{\hat{\Sigma}}$ is unitary and maps the non-contact modal frame to the contact modal frame; it is therefore a natural candidate for a gluing map between the modal bundles on the two sides of $\hat{\Sigma}$. The positive factor in Eq. (C3) contains the non-isometric part of the modal mismatch, while $Q_{\hat{\Sigma}}$ contains the phase and internal rotation information that can contribute to topology.

Under independent unitary changes of the contact and non-contact modal frames, $O_{\hat{\Sigma}}$ and its polar factor transform covariantly by left and right unitary multiplication. Therefore, $Q_{\hat{\Sigma}}$ should not be interpreted as a gauge-independent matrix at a fixed point. Its topological meaning is instead associated with its homotopy class, its holonomy along closed paths, or its combination with the Berry connections defined on the adjacent regions.

**C.3. Gluing of the modal bundles**

When $O_{\hat{\Sigma}}$ is invertible and $Q_{\hat{\Sigma}}$ varies continuously along $\hat{\Sigma}$, the data consisting of the restrictions of $M_{\mathrm{m}}$ to the adjacent regions and the interface map $Q_{\hat{\Sigma}}$ define a piecewise-smooth modal bundle with interface gluing. In this situation, the discontinuous band structure can be regarded as a pair of smooth band bundles together with a transition map on their common boundary.

For a complex single band, $Q_{\hat{\Sigma}}$ takes values in $U(1)$. If $\ell$ is a closed curve contained in $\hat{\Sigma}$ and $Q_{\hat{\Sigma}}$ is continuous along $\ell$, its winding is

$$\mathrm{wind}(\det Q_{\hat{\Sigma}}) = \frac{1}{2\pi i}\oint_{l} d\log(\det Q_{\hat{\Sigma}}) \qquad (C5)$$

Equation (C5) is an integer when the gluing map is continuous and remains invertible along $\ell$. Its value is invariant under gauge transformations whose determinant has zero winding along the curve, in particular transformations that extend over a surface in each adjacent region bounded by the curve. For several coupled bands, a gauge-covariant quantity is obtained from a closed Wilson loop formed by the path-ordered Berry transports in the two adjacent regions and the interface map $Q_{\hat{\Sigma}}$. Here ℓ is a closed curve and $Q_{\hat{\Sigma}}$ is continuous and invertible along it; admissible gauge changes must extend smoothly into the adjacent regions.

The important point is that the polar factor of the overlap matrix defines topological gluing data only after the following conditions are satisfied [50,51,77]:

• the fixed-rank modal subspace is isolated on both sides of $\hat{\Sigma}$;

• $O_{\hat{\Sigma}}$ is invertible on the considered interface;

• $Q_{\hat{\Sigma}}$ is continuous along the relevant part of $\hat{\Sigma}$;

• the resulting winding or holonomy is invariant under admissible gauge transformations.

Without these conditions, $O_{\hat{\Sigma}}$ remains a measure of modal mismatch rather than a transition function of a vector bundle.

**C.4. Rank-changing discontinuities and relative topology**

Within a fixed spectral window, contact constraints can change the number of eigenvalues in the selected isolated cluster, as illustrated by the change in the number of low-frequency bands between the non-contact and contacted cases. In such situations, the ranks of the corresponding spectral projections on the two sides of $\hat{\Sigma}$ need not be equal. The overlap matrix $O_{\hat{\Sigma}}$ is then rectangular or rank deficient, and an ordinary vector-bundle gluing construction is no longer applicable.

This rank-changing case is precisely where relative topology becomes relevant. More precisely, for a fixed spectral window $I$ that remains separated from the rest of the spectrum on each side, let $P_{\pm} = 1_{I(A_{\pm})}$ be the corresponding spectral projections. The appropriate object is not a single globally defined bundle over $P$, but the relative data consisting of the modal bundles on the smooth regions and the defect of their interface identification. In other words, this defect can be described using the pair of spectral projections associated with the two contact configurations. If the pair of projections is Fredholm, its index provides a standard integer-valued invariant [77].

Accordingly, the discontinuity contribution may be formulated as a relative class associated with the restrictions of $M_{\mathrm{m}}$ to the adjacent regions and the interface identification represented by $Q_{\hat{\Sigma}}$, or, in the rank-changing case, with the corresponding pair of spectral projections. This formulation is more general than assigning a separate Chern number to each individual eigenvector. It also accommodates the appearance or disappearance of admissible modes at the contact transition.

The bulk contribution is obtained from the Berry curvature on each smooth partition, while the interface contribution measures the obstruction to identifying the two modal bundles across $\hat{\Sigma}$. Such a relative class can change when the interface map becomes singular, the bulk spectral gap closes, or the relevant symmetry class changes.

**C.5. Relation to the stiffness-based transition matrix**

On a common finite-dimensional, mass-orthonormal modal subspace with basis matrix $U$, let $\widetilde{K}_{nc} = U^{\dagger} K_{nc} U$ and $\widetilde{K}_{cross} = U^{\dagger} K_{c} U$, where $K_{nc}$ and $K_{c}$ denote the non-contact and contacted stiffness matrices. Assuming positive-definite square roots on this subspace, define $R = \left(\widetilde{K}_{nc}^{1/2} + \widetilde{K}_{cross}^{1/2}\right)^{-1} \times \left(\widetilde{K}_{nc}^{1/2} - \widetilde{K}_{cross}^{1/2}\right)$. Its non-diagonal elements describe the coupling of modes induced by the contact configuration.

This provides a direct mechanical interpretation of modal conversion and energy redistribution at the discontinuity.

However, $R$ is generally not a unitary transition function. It depends on the stiffness representation, the modal normalization, and the chosen truncation of the modal basis. In particular, $Tr\ log\ R$ is not automatically gauge invariant, integer valued, or well defined in the infinite-dimensional limit. It should therefore not be identified directly with a topological invariant.

When $R$ is nonsingular on a common finite-dimensional modal subspace, its unitary polar factor $R\left(R^{\dagger}R\right)^{-1/2}$ may be compared with $Q_{\hat{\Sigma}}$. Such a comparison separates the geometric gluing of the modal bundles from the dynamical mode conversion caused by the contact stiffness. The former can carry topological information, whereas the latter determines the strength and structure of the mechanical scattering process.

Thus, the role of the two matrices is distinct: $O_{\hat{\Sigma}}$ characterizes the geometric overlap of modal subspaces, $Q_{\hat{\Sigma}}$ represents the unitary part of the interface gluing, and $R$ represents contact-induced modal transition and scattering. A nonzero off-diagonal element of $R$, or of $O_{\hat{\Sigma}}$, is not sufficient to demonstrate a topological phase transition.

**C.6. Spectral flow and the Dirac construction**

The contact discontinuity connects two different spectral families. It should therefore be described as a spectral transition rather than automatically as a spectral flow. Spectral flow requires a specified continuous family of self-adjoint Fredholm operators or an equivalent interface evolution. A discontinuous jump alone does not define a unique spectral flow because different regularizations of the jump can produce different intermediate spectra.

The Dirac-type construction discussed in Appendix D is used here only as a formal auxiliary factorization, denoted schematically by $D^2 = L + V$. If a graded elliptic Fredholm realization $D^+$ is constructed with specified domain and boundary/interface conditions, its index may be related to the topology of the modal bundles. In that conditional setting, a schematic decomposition has the form

$$ind\ D^{+} = T_{bulk} + T_{boundary} + T_{\hat{\Sigma}} \qquad (C6)$$

where, schematically, $T_{bulk}$ is determined by the characteristic forms constructed from $\mathcal{A}_m(\boldsymbol{P})$ and $\Omega_m(\boldsymbol{P})$. $T_{boundary}$ contains the boundary spectral-asymmetry correction, and $T_{\hat{\Sigma}}$ represents the relative interface class.

The boundary term in Eq. (C6) cannot in general be replaced by a Chern-character integral alone. For an APS-type formulation, the boundary operator, its spectral projection, and the associated η-invariant must be specified. Similarly, the interface contribution must be defined from the gluing map or from a Fredholm pair of spectral projections. These analytic conditions are additional requirements for applying an APS-type index theorem to the graded operator and are not consequences of the factorization $D^2 = L + V$ alone. Because Appendix D does not establish such a Fredholm realization for the contact operator, Eq. (C6) is only a conditional template and is not used as a result of this manuscript.

### *Appendix D: Construct the Dirac operator from isotropic elastic solid equation*

To obtain $\mathcal{D}\boldsymbol{\Psi} = \omega\boldsymbol{\Psi}$, Deymier and Runge introduced a pseudospin degree of freedom to describe the polarization of elastic waves, enabling the Dirac factorization of the two-dimensional elastic plate equations[78]. By dividing the elastic wave field into longitudinal waves and transverse waves, they obtained the following form of the Dirac operator:

$$\mathcal{D}_0 = \begin{pmatrix} & & \sum_{i=x,y,z} \theta_i \partial_i \\ & \sum_{i=x,y,z} \theta_i \partial_i & \\ \sum_{i=x,y,z} \theta_i \partial_i & & \end{pmatrix} + \begin{pmatrix} \omega/c_p & & \\ & \omega/c_s & \\ & & \omega/c_s \end{pmatrix}$$

Where $\theta_i$ is Pauli matrix, $c_p = \sqrt{(\kappa + 2\lambda)/\rho}$, $c_s = \sqrt{\lambda/\rho}$ are the wave speeds of longitudinal waves and transverse waves respectively. This satisfies ${\mathcal{D}_0}^2 = \mathcal{L}$. But considering the contact, we should further construct a Dirac operator $\mathcal{D} = \mathcal{D}_0 + \mathcal{W}$ makes $\mathcal{D}^2 = \mathcal{L} + \mathcal{V}$. This requires that $\mathcal{W}$ satisfy the super-symmetry algebraic relation:

$$\{\mathcal{D}_0, \mathcal{W}\} + \mathcal{W}^2 = \mathcal{V}$$

Choose $\mathcal{W} = \gamma\sqrt{\mathcal{V}}$, where $\gamma$ is a Clifford generator that is anti-commutative with $\mathcal{D}_0$. We can verify that

$$\mathcal{D}^2 = {\mathcal{D}_0}^2 + \mathcal{D}_0\gamma\sqrt{\mathcal{V}} + \gamma\sqrt{\mathcal{V}}\mathcal{D}_0 + (\gamma\sqrt{\mathcal{V}})^2$$

Using the inverse relation $\{\mathcal{D}_0, \gamma\} = 0, \gamma^2 = 1$, yield

$$\mathcal{D}^2 = {\mathcal{D}_0}^2 + \mathcal{V} = \mathcal{L} + \mathcal{V}$$

In this way, we have incorporated the contact force into the Dirac operator. In this manuscript, the Dirac factorization is used only as the auxiliary construction reported in ref [78], it cannot directly extent to the contact operator.

### ***Appendix E*: *Numerical implementation details***

We use ANSYS Mechanical software to solve equation (3) with both transient and static case in order to have a better efficiency, and use Comsol Multiphysics software with "Solid mechanics" module to solve equation (4) because of its convenience to process a parameter sweep and output the data of the band structure. The bulk material is set to density $1410\ \mathrm{kg/m^3}$, Young's modules 2.8 GPa, Poisson's ratio 0.4. The contact algorithm employs the augmented Lagrangian method. The large geometric deformation was ignored. Hexahedron structured mesh is used to improve the accuracy. Four mesh configurations with increasing accuracy with number of cells 2178,5224,10802,24820 for modal analysis are tested. The third configuration is chosen.

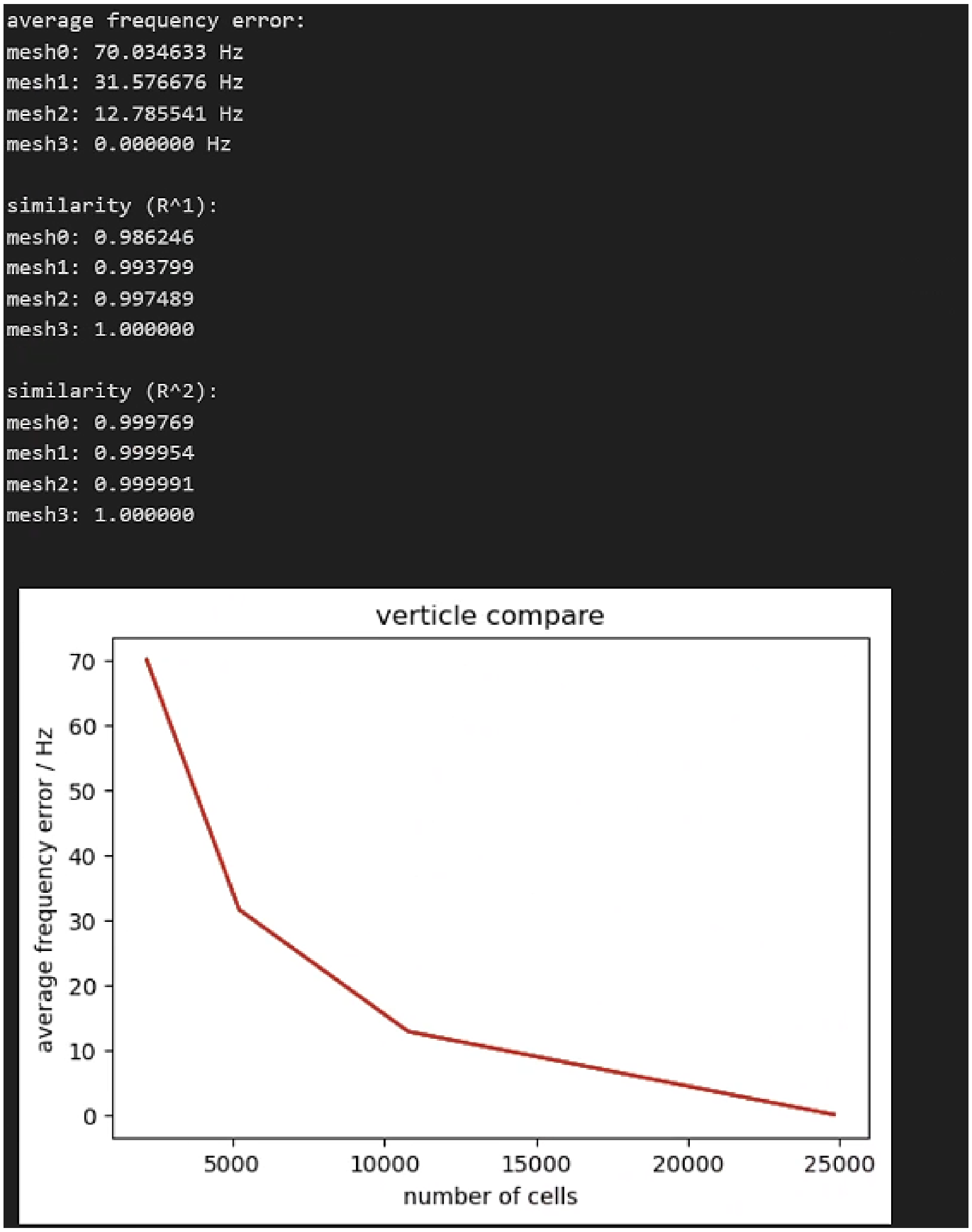

*Appendix F: Model validation of Near-DS perturbation*

We use the following case to verify the availability of the calculation method mentioned in Sec. III.C for sound velocity. Considering a 1.2-meter-long solid rod with $\rho = 1410\,\mathrm{kg}/m^3$, $E = 2.8\,\mathrm{GPa}$, the boundary conditions are consistent with those in Sec. III.C. The excitation mainly stimulates the pressure wave, it passes though the rod using 0.855 ms (As shown in the figure below, the red curve represents the average pressure of the output interface), yield its speed $1403.51\ \mathrm{m/s}$. Compared with the analytic value $\sqrt{E/\rho} = 1409.19\ \mathrm{m/s}$, an error of 0.4% is given.

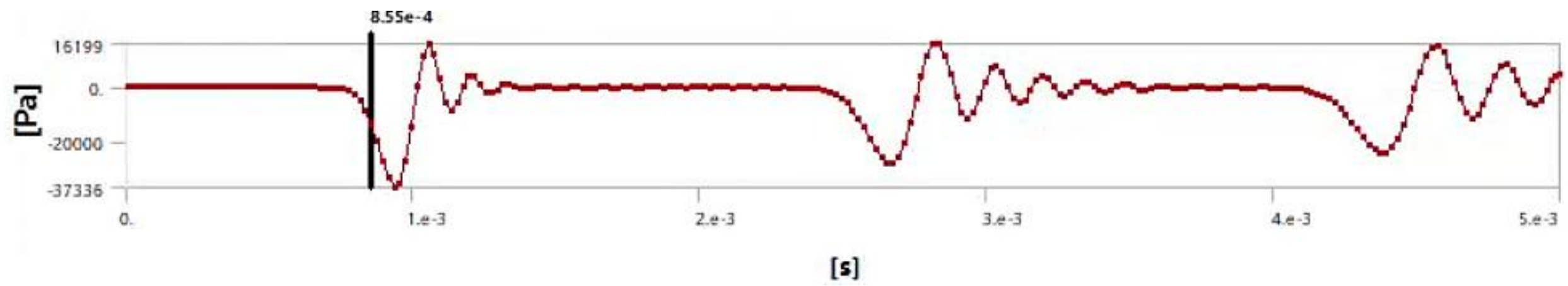

*Figure S1 Probe area-mean-stress of the verification case*

## *Appendix G: Density of states*

$D(\omega) = 2\sum_m \int_0^{\pi/a} \frac{\mathrm{d}k_x}{\omega_m'(k_x)}$ is the density of states (DOS), it determines the transport properties of sound and heat in DPMMs with the group velocity. It can be also tuned by a small strain across different contact states.

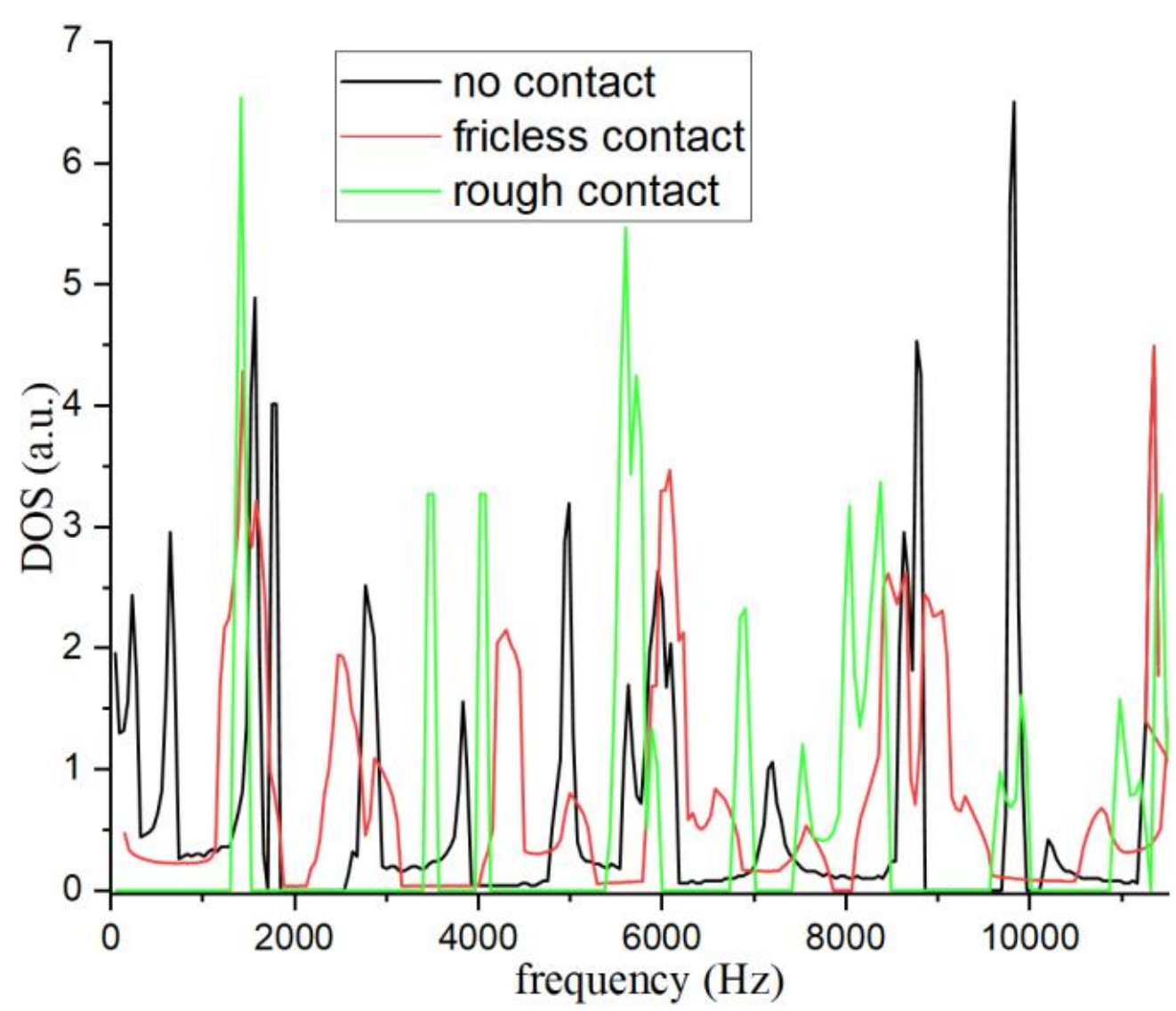

*Figure S2: Phonon density of states in the three studied cases.*

**References**

1. Kadic, M., Bückmann, T., Schittny, R. & Wegener, M. Metamaterials beyond electromagnetism. *Reports on Progress in Physics* **76**, 126501 (2013).

2. Yu, X., Zhou, J., Liang, H., Jiang, Z. & Wu, L. Mechanical metamaterials associated with

stiffness, rigidity and compressibility: A brief review. *Progress in Materials Science* **94**, 114–173 (2018).

3. Surjadi, J. U. *et al.* Mechanical Metamaterials and Their Engineering Applications. *Advanced Engineering Materials* **21**, 1800864 (2019).

4. Bertoldi, K., Vitelli, V., Christensen, J. & van Hecke, M. Flexible mechanical metamaterials. *Nature Reviews Materials* **2**, 17066 (2017).

5. Zheng, X. *et al.* Ultralight, ultrastiff mechanical metamaterials. *Science* **344**, 1373–1377 (2014).

6. Babaee, S. *et al.* 3D Soft Metamaterials with Negative Poisson's Ratio. *Advanced Materials* **25**, 5044–5049 (2013).

7. Yang, N. *et al.* Mechanical Metamaterials with Discontinuous and Tension/Compression-Dependent Positive/Negative Poisson's Ratio. *Advanced Engineering Materials* **24**, 1–9 (2022).

8. Rong, L., Yifeng, Z., Haiwen, C., Yuxin, T. & Minfang, C. Enhancing performance of sandwich panel with three-dimensional orthogonal accordion cores. *International Journal of Mechanical Sciences* **285**, 109787 (2025).

9. Rueger, Z. & Lakes, R. S. Cosserat elasticity of negative Poisson's ratio foam: experiment. *Smart Materials and Structures* **25**, 54004 (2016).

10. Jiang, H. EML webinar overview: Origami-based metamaterials. *Extreme Mechanics Letters* **50**, 101543 (2022).

11. Ji, J. C., Luo, Q. & Ye, K. Vibration control based metamaterials and origami structures: A state-of-the-art review. *Mechanical Systems and Signal Processing* **161**, (2021).

12. Tan, X. *et al.* Real-time tunable negative stiffness mechanical metamaterial. *Extreme Mechanics Letters* **41**, 100990 (2020).

13. Li, H., Li, Y. & Li, J. Negative stiffness devices for vibration isolation applications: A review. *Advances in Structural Engineering* **23**, 1739–1755 (2020).

14. Corsi, M., Bagassi, S., Moruzzi, M. C. & Weigand, F. Additively manufactured negative stiffness structures for shock absorber applications. *Mechanics of Advanced Materials and Structures* **29**, 999–1010 (2022).

15. Zhang, K., Qi, L., Zhao, P., Zhao, C. & Deng, Z. Buckling induced negative stiffness mechanical metamaterial for bandgap tuning. *Composite Structures* **304**, 116421 (2023).

16. Bergman, D. J. & Stroud, D. Physical Properties of Macroscopically Inhomogeneous Media. in (eds. EHRENREICH, H. & TURNBULL, D.) vol. 46 147–269 (Academic Press, 1992).

17. Ho, P. P. & Alfano, R. R. Optical Kerr effect in liquids. *Phys. Rev. A* **20**, 2170–2187 (1979).

18. Righini, R. Ultrafast Optical Kerr Effect in Liquids and Solids. *Science* **262**, 1386–1390 (1993).

19. Kouwenhoven, L. & Glazman, L. Revival of the Kondo effect. *Physics World* **14**, 33 (2001).

20. Hewson, A. C. *The Kondo Problem to Heavy Fermions*. (Cambridge University Press, 1993).

21. Castañeda, P. P. Nonlinear Composite Materials: Effective Constitutive Behavior and Microstructure Evolution. in *Continuum Micromechanics* (ed. Suquet, P.) 131–195 (Springer Vienna, Vienna, 1997). doi:10.1007/978-3-7091-2662-2_3.

22. Maddouri, K., Triki, H., Wang, N. & Zhou, Q. Nonlinear tunneling of chirped similaritons in non-centrosymmetric waveguides with quadratic-cubic nonlinearity. *Physics Letters A* **512**, 129584 (2024).

23. Ma, G., Xiao, M. & Chan, C. T. Topological phases in acoustic and mechanical systems. *Nature Reviews Physics* **1**, 281–294 (2019).

24. Heiskanen, S., Puurtinen, T. A. & Maasilta, I. J. Controlling thermal conductance using three-dimensional phononic crystals. *APL Materials* **9**, 081108 (2021).

25. Gao, Y., Kang, X. & Li, B. Programmable mechanical metamaterials with tunable Poisson's ratio and morphable stiffness. *Composites Part B: Engineering* **292**, 112089 (2025).

26. Liang, X., Fu, H. & Crosby, A. J. Phase-transforming metamaterial with magnetic interactions. *Proceedings of the National Academy of Sciences* **119**, e2118161119 (2022).

27. Nicolaou, Z. G. & Motter, A. E. Mechanical metamaterials with negative compressibility transitions. *Nature Materials* **11**, 608–613 (2012).

28. Kim, E. Tunable frequency bandgaps in elastic metamaterials with internal contacts. *INTER-NOISE* 1–4.

29. Lazarides, N. & Tsironis, G. P. Superconducting metamaterials. *Physics Reports* **752**, 1–67 (2018).

30. Li, F., Chong, C., Yang, J., Kevrekidis, P. G. & Daraio, C. Wave transmission in time- and space-variant helicoidal phononic crystals. *Physical review. E, Statistical, nonlinear, and soft matter physics* **90 5–1**, 53201 (2014).

31. Patil, G. U., Cui, S. & Matlack, K. H. Leveraging nonlinear wave mixing in rough contacts-based phononic diodes for tunable nonreciprocal waves. *Extreme Mechanics Letters* **55**, 101821 (2022).

32. Schwarz, D., Felsch, G., Tauber, F., Schiller, S. & Slesarenko, V. Exploiting self-contact in mechanical metamaterials for new discrete functionalities. *Materials & Design* **236**, 112468 (2023).

33. He, J., Liang, T., Wang, Y., Xia, L. & Xiong, Y. Quasi-zero stiffness metamaterials with programmable multistability for multi-scenario applications. *Applied Materials Today* **46**, 102864 (2025).

34. Wu, L. & Pasini, D. Zero modes activation to reconcile floppiness, rigidity, and multistability into an all-in-one class of reprogrammable metamaterials. *Nature Communications* **15**, 3087 (2024).

35. Simpson, D. J. W. Border-collision bifurcations from stable fixed points to any number of coexisting chaotic attractors. *Journal of Difference Equations and Applications* **30**, 90–110 (2024).

36. Nusse, H. E. & Yorke, J. A. Border-collision bifurcations including "period two to period three" for piecewise smooth systems. *Physica D: Nonlinear Phenomena* **57**, 39–57 (1992).

37. Dutta, M., Nusse, H. E., Ott, E., Yorke, J. A. & Yuan, G. Multiple attractor bifurcations: A source of unpredictability in piecewise smooth systems. *Physical Review Letters* **83**, 4281–4284 (1999).

38. Champneys, A. & Bernardo, M. *Piecewise Smooth Dynamical Systems*. *Scholarpedia* vol. 3 (2008).

39. Glendinning, P. A. & Simpson, D. J. W. Chaos in the border-collision normal form: A computer-assisted proof using induced maps and invariant expanding cones. *Applied Mathematics and*

*Computation* **434**, 1–21 (2022).

40. Zhang, H. W., Liao, A. H., Xie, Z. Q., Chen, B. S. & Wang, H. Some advances in mathematical programming method for numerical simulation of contact problems. *Solid Mechanics and its Applications* **3**, 33–55 (2007).

41. Kunze, M. *Non-Smooth Dynamical Systems*. (springer, 2000).

42. Strömberg, N., Johansson, L. & Klarbring, A. A Generalised Standard Model for Contact, Friction and Wear. in *Contact Mechanics* (eds. Raous, M., Jean, M. & Moreau, J. J.) 327–334 (Springer US, Boston, MA, 1995). doi:10.1007/978-1-4615-1983-6_44.

43. Luo, A. C. J. A theory for non-smooth dynamic systems on the connectable domains. *Communications in Nonlinear Science and Numerical Simulation* **10**, 1–55 (2005).

44. Billups, S. C. & Murty, K. G. Complementarity problems. *Journal of Computational and Applied Mathematics* **124**, 303–318 (2000).

45. Júdice, J. J., Sherali, H. D. & Ribeiro, I. M. The eigenvalue complementarity problem. *Computational Optimization and Applications* **37**, 139–156 (2007).

46. Reed, M. & Simon, B. *Methods of Modern Mathematical Physics Volume 3: Scattering Theory*. (Academic Press, 1979).

47. Kuhlmann, M. & Stöckler, M. *Quantum Field Theory*. (Westview Press, 2018). doi:10.1007/978-3-319-78356-7_6.

48. Gottlieb, D. & Shu, C.-W. On the Gibbs Phenomenon and Its Resolution. *SIAM Review* **39**, 644–668 (1997).

49. Legrand, M., Junca, S. & Heng, S. Nonsmooth modal analysis of a N-degree-of-freedom system undergoing a purely elastic impact law. *Communications in Nonlinear Science and Numerical Simulation* **45**, 190–219 (2017).

50. Hislop, P. D. & Sigal, I. M. Self-Adjointness: Part 2. The Kato-Rellich Theorem. in *Introduction to Spectral Theory: With Applications to Schrödinger Operators* 131–138 (Springer New York, New York, NY, 1996). doi:10.1007/978-1-4612-0741-2_13.

51. Tosio Kato. *Perturbation Theory for Linear Operators*. (1995).

52. Xie, B. *et al.* Dirac points and the transition towards Weyl points in three-dimensional sonic crystals. *Light: Science and Applications* **9**, (2020).

53. Tarruell, L., Greif, D., Uehlinger, T., Jotzu, G. & Esslinger, T. Creating, moving and merging Dirac points with a Fermi gas in a tunable honeycomb lattice. *Nature* **483**, 302–305 (2012).

54. Herrera, M. A. J. & Bercioux, D. Tunable Dirac points in a two-dimensional non-symmorphic wallpaper group lattice. *Communications Physics* **6**, 42 (2023).

55. Armitage, N. P., Mele, E. J. & Vishwanath, A. Weyl and Dirac semimetals in three-dimensional solids. *Reviews of Modern Physics* **90**, 15001 (2018).

56. Ma, J. Z. *et al.* Observation of a singular Weyl point surrounded by charged nodal walls in PtGa. *Nature Communications* **12**, 1–8 (2021).

57. Liu, Q. B., Qian, Y., Fu, H. H. & Wang, Z. Symmetry-enforced Weyl phonons. *npj Computational Materials* **6**, 1–6 (2020).

58. Yang, Y. Phononic nodal point in two dimensions: A mini-review. *Frontiers in Physics* **10**, 8–13 (2022).

59. Wang, X. *et al.* Topological nodal line phonons: Recent advances in materials realization. *Applied Physics Reviews* **9**, (2022).

60. Park, H., Gao, W., Zhang, X. & Oh, S. S. Nodal lines in momentum space: Topological invariants and recent realizations in photonic and other systems. *Nanophotonics* **11**, 2779–2801 (2022).

61. Asano, K. & Hotta, C. Designing Dirac points in two-dimensional lattices. *Physical Review B - Condensed Matter and Materials Physics* **83**, 1–14 (2011).

62. Dubček, T. *et al.* Weyl Points in Three-Dimensional Optical Lattices: Synthetic Magnetic Monopoles in Momentum Space. *Physical Review Letters* **114**, (2015).

63. Tempelman, J. R., Vakakis, A. F. & Matlack, K. H. Wavenumber Scattering and Inter-band Targeted Energy Transfer in Phononic Lattices with Local Vibro-Impact Nonlinearities.

64. Grinberg, I. & Matlack, K. H. Nonlinear elastic wave propagation in a phononic material with periodic solid–solid contact interface. *Wave Motion* **93**, 102466 (2019).

65. Anufriev, R. & Nomura, M. Reduction of thermal conductance by coherent phonon scattering in two-dimensional phononic crystals of different lattice types. *Phys. Rev. B* **93**, 45410 (2016).

66. Klitsner, T. & Pohl, R. O. Phonon scattering at silicon crystal surfaces. *Phys. Rev. B* **36**, 6551–6565 (1987).

67. MITRA, S. K. & RAKSHIT, H. Refraction of Light Waves by Electrons. *Nature* **123**, 796–797 (1929).

68. Lauterborn, W. & Holzfuss, J. ACOUSTIC CHAOS. *INTERNATIONAL JOURNAL OF BIFURCATION AND CHAOS* **1**, 13–26 (1991).

69. Arecchi, F. T. Instabilities and Chaos in Optics. *Physica Scripta* **T23**, 160–164 (1988).

70. Hosham, H. A. & Alharthi, T. N. Bifurcation and chaos in simple discontinuous systems separated by a hypersurface. *AIMS Mathematics* **9**, 17025–17038 (2024).

71. Zhusubaliyev, Z. T. & Mosekilde, E. *Bifurcations and Chaos in Piecewise-Smooth Dynamical Systems*. (WORLD SCIENTIFIC, 2003). doi:10.1142/5313.

72. Alexander, J. C., Yorke, J. A., You, Z., & Kan, I. Riddled Basins. *International Journal of Bifurcation and Chaos* **02**, 795–812 (1992).

73. Hua, J., Zihao, Z. & Xin, W. Discontinuous bifurcations of coexisting attractors for a gear transmission system. **51**, 1145–1155 (2021).

74. Banerjee, S., Yorke, J. A. & Grebogi, C. Robust chaos. *Physical Review Letters* **80**, 3049–3052 (1998).

75. Ghosh, I., McLachlan, R. I. & Simpson, D. J. W. Robust chaos in orientation-reversing and non-invertible two-dimensional piecewise-linear maps. 1–27 (2023).

76. Zee, L. & Sternberg, E. Ordinary and strong ellipticity in the equilibrium theory of incompressible hyperelastic solids. *Arch. Ration. Mech. Anal.* **83**, 53–90 (1983).

77. Avron, J., Seiler, R. & Simon, B. The Index of a Pair of Projections. *Journal of Functional Analysis* **120**, 220–237 (1994).

78. Deymier, P. A. & Runge, K. Revealing topological attributes of stiff plates by Dirac factorization of their 2D elastic wave equation. *Applied Physics Letters* **120**, (2022).